\documentclass[12pt]{article}%
\usepackage[applemac]{inputenc}
\usepackage{amsmath,amssymb,fullpage}
\usepackage{amsfonts}
\usepackage{amsmath,amssymb}
\usepackage[applemac]{inputenc}
\usepackage{amsmath,amssymb,fullpage}
\usepackage{color}
\usepackage{color, colortbl}
\usepackage{amsmath}
\usepackage{amssymb}
\usepackage{graphicx}
\usepackage{tikz}
\usepackage{geometry}
\usetikzlibrary{arrows}
\usepackage{amsfonts}
\usepackage{amsmath,amssymb}
\usepackage[applemac]{inputenc}
\usepackage{amsmath,amssymb,fullpage}
\usepackage{color}
\usepackage{amsmath}
\usepackage{amssymb}
\usepackage{graphicx}
\usepackage{tikz}
\usepackage{graphicx}
\graphicspath{ {./images/} }
\usepackage{amsmath}
\usepackage{algorithm}
\usepackage{float}
\usepackage[round]{natbib}
\usepackage{algpseudocode}
\usepackage[section]{placeins}
\usepackage{multirow}
\usepackage{caption}
\usepackage{subcaption}
\usepackage{bbm, dsfont}
\usepackage{booktabs}

\usepackage[final]{changes}

\newtheorem{theorem}{Theorem}[section]

\newtheorem{proposition}[theorem]{Proposition}
\newtheorem{corollary}[theorem]{Corollary}

\newtheorem{problem}[theorem]{Problem}
\newtheorem{remark}[theorem]{Remark}

\usepackage{tikz}
\usetikzlibrary{shapes.geometric, arrows}

\tikzstyle{input} = [circle, minimum width=1cm, text centered, draw=black, fill=green!20]

\tikzstyle{output} = [circle, minimum width=1cm, text centered, draw=black, fill=blue!20]

\tikzstyle{lstm} = [rectangle, rounded corners, minimum width=2cm, minimum height=1cm,text centered, draw=black, fill=red!20]

 \tikzstyle{lin} = [rectangle, minimum width=2cm, minimum height=1cm,text centered, draw=black, fill=orange!20]

 \tikzstyle{act} = [ellipse, minimum width=2cm, minimum height=1cm,text centered, draw=black, fill=yellow!20]

\tikzstyle{dot} = [rectangle, minimum width=2cm, minimum height=1cm,text centered]

\tikzstyle{arrow} = [thick,->,>=stealth]
\tikzstyle{map} = [thick, dashed,->,>=stealth]

\usetikzlibrary{arrows,shapes,automata,petri}
\newenvironment{myenumerate}{

\begin{enumerate}}{\end{enumerate}}
\newcommand{\dproof}{\noindent {Proof.} \quad}
\newcommand{\fproof}{\hfill $\square$ \bigskip}

\numberwithin{equation}{section}

\def\RB{\mathbb{R}}

\definecolor{LightCyan}{rgb}{0.88,1,1}

\def\1B{\text{1\!\!I}}

\begin{document}

\title{Deep learning for quadratic hedging in incomplete jump market}
\author{Nacira Agram$^{1}$, Bernt \O ksendal$^{2}$ \& Jan Rems$^{3}$}
\date{\today}\maketitle

\footnotetext[1]{Department of Mathematics, KTH Royal Institute of Technology 100 44, Stockholm, Sweden. \newline
Email: nacira@kth.se. Work supported by the Swedish Research Council grant (2020-04697).}

\footnotetext[2]{
Department of Mathematics, University of Oslo, Oslo, Norway. \newline
Email: oksendal@math.uio.no.}

\footnotetext[3]{Department of Mathematics, University of Ljubljana, Ljubljana, Slovenia. \newline Email: jan.rems@fmf.uni-lj.si}

\begin{abstract}
We propose a deep learning approach to study the minimal variance pricing and hedging problem in an incomplete jump diffusion market. It is based upon a rigorous stochastic calculus derivation of the optimal hedging portfolio, optimal option price, and the corresponding equivalent martingale measure through the means of the Stackelberg game approach. A deep learning algorithm based on the combination of the feedforward and LSTM neural networks is tested on three different market models, two of which are incomplete. In contrast, the complete market Black-Scholes model serves as a benchmark for the algorithm's performance. The results that indicate the algorithm's good performance are presented and discussed. 

In particular, we apply our results to the special incomplete market model studied by Merton and give a detailed comparison between our results based on the minimal variance principle and the results obtained by Merton based on a different pricing principle. Using deep learning, we find that the minimal variance principle leads to typically higher option prices than those deduced from the Merton principle. On the other hand, the minimal variance principle leads to lower losses than the Merton principle.

\end{abstract}



\vskip 0.5cm




\textbf{Keywords :} Option pricing; Incomplete market; Equivalent martingale measure; Merton model; Deep learning; LSTM.
\section{Introduction}
Many jump diffusion models have been extensively studied in the past. Of particular interest to us is the Merton model proposed by \cite{M}, where the jump distribution is log-normal. Merton derived a closed-form solution to the hedging problem for the European option, assuming diversifiabilty of the jump component. There are also other popular jump diffusion models. \cite{kou2002jump} proposed an alternative approach where double exponential jump distribution is assumed. The variance gamma model proposed in \citep{madan1990variance} also supports the inclusion of jumps.

Regarding the use of machine learning in stochastic control theory, there have been a lot of new developments in the past few years. With new techniques, it has now become possible to solve problems that do not obtain closed-form solutions numerically. In \citep*{han2016deep}, a deep learning approach for solving stochastic control problems was proposed. The idea is that one discretizes time and, at each timestep, approximates the control with a feed-forward neural network. In the papers \citep*{buehler2019deep,carbonneau2021equal}, the authors extend these results to reinforcement learning algorithms for hedging in high dimensions. Methods proposed in \citep*{han2016deep} also provide a strong tool for finding the numerical solutions to high dimensional partial differential equations. In \added{a seminal paper \citep*{ han2018} and later on in} \citep*{beck2019machine, chan2019machine}, the authors transform PDEs to backward stochastic differential equations and use deep learning to solve the associated stochastic control problem. 

Since control problems related to hedging rely heavily on time series dynamics, it is convenient to consider recurrent neural networks (RNN) instead of feed-forward ones. These types of networks were proposed in \citep*{rumelhart1986learning}. The recurring nature of the network can contain cycles, which makes them suitable for problems that have sequential inputs. A particular case of RNN called long short-term memory (LSTM) was first introduced in \citep*{HJ} and became very prominent in the last few years for time series modelling. The main advantage over RNN is LSTM's ability to provide a memory structure that can store information over a large number of timesteps. It has been heavily used in financial time series prediction; for instance, see \citep*{cao2019financial, selvin2017stock, siami2018comparison}. One of the main advantages of LSTM networks is that they can also be used for non-Markovian models. In particular, Han et al. used them in \citeyearpar{han2021recurrent} to apply the methods proposed in \citep*{han2016deep} to stochastic control problems with delay. Deep learning for mean-field stochastic control has been studied by \cite{agram2020deep} where the initial state was a part of the control process and then extended to mean-field systems with delay in \citep*{agram2022deep}. \cite{lai2022data} also used LSTM networks to develop a data-driven approach for options market making. 

Since our main concern is hedging in an incomplete market, we also refer to \citeyearpar{FE}, where Fecamp et al. present several deep learning algorithms for discrete-time hedging, where the sources of incompleteness are illiquidity, non-tradable risk factors, and transaction costs. They propose a modified global LSTM approach where a single LSTM net is used that takes the whole time series as an input. The above approach was adopted in \citep*{gao2021inflexible}, where they study the double exponential jump distribution model proposed in \citep*{kou2002jump}. 

Furthermore, we do not consider the calibration of SDEs to fit market data in this paper. For promising approaches, one can check \citep*{boursin2022deep,remlinger2022conditional} where generative adversarial networks are used. Another approach could feature neural SDEs. A good survey is \citep*{gierjatowicz2020robust}.

In what follows, let us describe our purpose in the present paper.\\
We consider a financial market where the underlying price of a risky asset follows a jump diffusion process 
\begin{align}\label{S}
dS(t) & =S(t^{-})\left[  \alpha(t)dt+\sigma(t)dB(t)+\int_{
\mathbb{R^*}
}\gamma(t,\zeta)\widetilde{N}(dt,d\zeta)\right],\quad S(0) >0,
\end{align}
with $\mathbb{R}^{*}=\mathbb{R}\setminus \{0\}$, and $S_{0}(t)=1$ for all $t$ denotes the value of a risk-free money market
account.

Suppose now that the interest rate in the risk free asset equals 0 and the time to maturity is finite and denoted by $T$. \\
For simplicity of notation, we will write $S(t)$ instead of $S(t^{-})$ in the following.\\
Here $\alpha(t) \in \mathbb{R}, \sigma(t)=(\sigma_1(t), ..., \sigma_m(t)) \in \mathbb{R}^{m},$ $\gamma(t,\zeta)=(\gamma_1(t,\zeta),..., \gamma_k(t,\zeta)) \in \mathbb{R}^{k}$, and $B(t)=(B_1(t), ... , B_m(t))^{'} \in \mathbb{R}^{m}$ and  
$\widetilde{N}(dt,d\zeta)=(\widetilde{N}_1(dt,d\zeta), ..., \widetilde{N}_k(dt,d\zeta))^{'} \in \mathbb{R}^{k}$ are independent Brownian motions and compensated Poisson random measures, respectively, defined on a complete filtered probability space $(\Omega,\mathcal{F},\mathbb{F}=\{\mathcal{F}_t\}_{t\geq 0},P)$, where $\mathbb{F}=\{\mathcal{F}_{t}\}_{t\geq0}$ is the filtration generated by
$\{B(s)\}_{s\leq t}$ and $\{N(s,\zeta)\}_{s\leq t}$. We are using the matrix notation, i.e.
\begin{align*}
 \sigma(t)dB(t):&=\sum_{i=1}^{m} \sigma_i(t)dB_i(t),\quad
 \gamma(t,\zeta)\widetilde{N}(dt,d\zeta)= \sum_{j=1}^{k}\gamma_j(t,\zeta)\widetilde{N}_j(dt,d\zeta),
\end{align*}
and we assume that 
\begin{equation}\label{1.2}
\gamma(t,\zeta)>-1 \quad \text{ and } \quad  \sigma^2(t) +\int_{
\mathbb{R^*}
} \gamma^2(t,\zeta) \nu(d\zeta) > 0, \; \text{ for a.a.} \; t, \text{ a.s.}
\end{equation}
The coefficients $\alpha,\sigma$ and $\gamma$ are supposed to be deterministic and bounded.\\

The results in this paper can easily be extended to an arbitrary number of risky assets, but since the features of incomplete markets we are dealing with, can be fully illustrated by one jump diffusion risky asset only, we will for simplicity concentrate on this case in the following. We emphasise however, that we deal with an arbitrary number $m$ of independent Brownian motions and an arbitrary number $k$ of independent Poisson random measures in the representation \eqref{S}.

Let $z\in%
\mathbb{R}
$ be an initial endowment and let $\pi(t)\in \mathbb{R}$ be a self-financing portfolio,
representing the fraction of the total wealth $X(t)=X_{z,\pi}(t)$ invested in
the risky asset at time $t$. 
We say that $\pi$ is admissible if in addition  $\pi$ is a predictable process in $L^{2}(dt \times dP)$. The set of admissible portfolios is denoted by $\mathcal{A}$.
We associate to an admissible portfolio $\pi$ the wealth dynamics 
\begin{align}\label{W}
dX(t) =X(t)\pi(t)\left[  \alpha(t)dt+\sigma(t)dB(t)+\int_{
\mathbb{R^*}
}\gamma(t,\zeta)\widetilde{N}(dt,d\zeta)\right], \quad X(0) =z.
\end{align}
Here and in the following, we write for notational simplicity $X(t)$ instead of $X(t^{-})$, in agreement with our convention for $S(t)$.

 \vskip 0.2cm

Fix $T$ be the time to maturity and let $F$ be a given $T$-claim, i.e. $F \in L^2(P)$ is an  $\mathcal{F}_{T}$-measurable random variable, representing the payoff at that time. Then for each initial wealth $z\in%
\mathbb{R}
$ and each portfolio $\pi \in\mathcal{A}$, we want to minimize the expected squared hedging error
\begin{align*}
 J(z,\pi)=E\left[  \frac{1}{2}\left(  X_{z,\pi}(T)-F\right)  ^{2}\right]  .   
\end{align*}
Then the problem we consider is the following.
\begin{problem} \label{mv}
Find the optimal initial endowment $\widehat{z}\in%
\mathbb{R}
$ and the optimal portfolio $\widehat{\pi}\in\mathcal{A}$,  such that%
\begin{equation*}
\underset{z,\pi}{\inf}J(z,\pi)=J(\widehat{z},\widehat{\pi}). %
\end{equation*}
\end{problem}
Heuristically, this means that we define the price of the option with payoff $F$ to be the initial endowment $\widehat{z}$ needed to get the terminal wealth $X(T)$ as close as possible to $F$ in quadratic mean by an admissible portfolio $\widehat{\pi}$.


We may regard the minimal variance problem (Problem \ref{mv}) as a Stackelberg game, in which the first player chooses the initial endowment $z$, followed by the second player choosing the optimal portfolio $\pi$ based on this initial endowment. Knowing this response $\pi=\pi_{z}$ from the follower, the first player chooses the initial endowment $\widehat{z}$ which leads to a response $\pi=\pi_{\widehat{z}}$ which is optimal, in the sense that $J(\widehat{z},\pi_{\widehat{z}}) \leq J(z,\pi)$ over all admissible pairs $(z,\pi)$. 

For a general (discontinuous) semimartingale market, this is already a known result; we refer, for example, to  \citep*{cerny} where the authors have proved that the optimal quadratic hedging strategy $\widehat{\pi}$ can first be characterized in a linear feedback form which is independent of the initial endowment $z$, and that $\widehat{z}$ can then be identified in a second step.
Therefore, the results of Section 2 of the current paper might be derived from already existing results in the literature \added{\citep*{cerny}} but we found it convenient for the reader to find explicit expressions of the optimal quadratic hedging problem for a jump diffusion market (L\'evy process). \\
\cite{S1} proves (under some conditions, including a non-arbitrage condition) that there exists a signed measure $\Tilde{P}$, called the variance-optimal measure, such that
\begin{align}\label{sw}
    \widehat{z}= E_{\Tilde{P}}[F].
\end{align} We consider in the present paper a jump diffusion market and we show that $\Tilde{P}$ is a positive measure and we find it explicitly.

In the next section, 
we prove the  existence of the optimal hedging strategy and the optimal initial endowment for jump diffusion market.\\
We show that the minimal variance price can be represented as the expected value of the option's (discounted) payoff under some equivalent martingale measures (which is given explicitly). The analysis concludes with looking at specific examples of jump diffusion market models.

Section 3 is devoted to the deep learning approach. Since the option price and the optimal hedging portfolio can not always be computed explicitly, or it is not feasible, we need to use some other numerical methods. We propose a deep learning algorithm that approximates option price and hedging portfolio. We propose a joint feed-forward and multilayered LSTM network that adopts the "online" approximation approach proposed in \citep*{han2016deep}, where a neural network is used at each timestep to predict the control process. 

First, we test this algorithm in the case of a complete market, namely in the case of the Black Scholes (BS) model. This is done in order to estimate its performance since both option price and hedging portfolio can be obtained explicitly in the BS case. 

\replaced{Then we apply it to models that assume an incomplete market. First, a continuous model where the underlying asset depends on multiple independent Brownian motions is used in order to study the algorithm's scalability to multi-dimensional inputs. The same is then also done for the jump diffusion Merton model.}{Then apply it to models that assume an incomplete market. First, a continuous model where the underlying asset depends on two independent Brownian motions is used, and then we study a Merton model.} 

We show the algorithm's success in all three models and discuss its performance. Since Merton's reasoning behind his hedging strategy is not consistent with the behaviour of financial markets, we also compare our results to those obtained by Merton and discuss how the approaches differ. In particular, we show how the minimal variance approach provides a safer hedging strategy compared to the one proposed by Merton. \added{Finally, the algorithm's performance is also tested for another jump diffusion model, namely the Kou model.}\\

\section{Optimal quadratic hedging portfolio/optimal initial endowment} \label{sec:optimal control}
In this section, we find the optimal strategy pair to Problem \ref{mv}. 
\subsection{Equivalent martingale measures (EMMs)}
Since EMMs play a crucial role in our discussion, we start this section by recalling that an important group of measures $Q\in\mathbb{M}$ can be described as follows (we refer to Chapter 1 in \citep*{OS} 
 for more details):\\
Let $\theta_{0}(t)$ and $\theta_{1}(t,\zeta)>-1$ be $\mathbb{F}$-predictable
processes such that
\begin{align}\label{emm}
\alpha(t)+\theta_{0}(t)\sigma(t)+\int_{
\mathbb{R^*}
}\theta_{1}(t,\zeta)\gamma(t,\zeta)\nu(d\zeta)=0\text{, } \quad t\geq0.
\end{align}
Define the local martingale $Z(t)=Z^{\theta_{0},\theta_{1}}(t)$, by%

\begin{align*}
dZ(t)  =Z(t)\left[  \theta_{0}(t)dB(t)+\int_{\mathbb{R^*}
}\theta_{1}(t,\zeta)\widetilde{N}(dt,d\zeta)\right], \quad 
Z(0) =1,
\end{align*}
i.e.,
\small
\begin{align*}
Z(t)  &  =\exp\left(  \int_{0}^{t}\theta_{0}(s)dB(s)-\frac{1}{2}\int_{0}%
^{t}\theta_{0}^{2}(s)ds\right.  +\int_{0}^{t}\int_{\mathbb{R^*}
}\left\{  \ln(1+\theta_{1}(s,\zeta))-\theta_{1}(s,\zeta)\right\}  \nu
(d\zeta)ds\nonumber\\
&  \left.  +\int_{0}^{t}\int_{\mathbb{R^*}
}\ln(1+\theta_{1}(s,\zeta))\widetilde{N}(ds,d\zeta)\right).
\end{align*}
A sufficient condition for $Z(\cdot)$ being a true martingale is
\begin{align}\label{KS}
    E\Big[\exp\Big(\frac{1}{2}\int_0^T \theta_0^2(s)ds + \int_0^T\int_{\mathbb{R^*}}\theta_1^2(s,\zeta) N(ds,d\zeta)\Big)\Big] < \infty.
\end{align}
For proof see \citep*{KaSh}.
Then the measure $Q^{\theta_{0},\theta_{1}}$ defined by%
\begin{align*} 
dQ^{\theta_{0},\theta_{1}}(\omega)=Z(T)dP(\omega)\text{ on }\mathcal{F}_{T}%
\end{align*}
is in $\mathbb{M}$.

\subsection{The optimal portfolio}
Assume as before that the wealth process $X(t)=X_{z, \pi}(t)$, corresponding to an initial wealth $z$ and a self-financing portfolio $\pi$, is given by
\begin{align*}
    dX(t)=X(t)\pi(t)\Big[\alpha(t) dt + \sigma(t) dB(t) + \int_{ \mathbb{R^*} }\gamma(t,\zeta)\widetilde{N}(dt,d\zeta)\Big],\quad
    X(0)=z.
\end{align*}
Let $F$ be a given $T$-claim representing the terminal payoff of the option. By the It\^o/martingale representation theorem for jump diffusions (see \citep*{L}) we can write $F=F(T)$, where the martingale $F(t):=E[F|\mathcal{F}_t],$ $ t \in [0,T]$ has the  It\^o-L\'evy representation
\begin{align*}
    dF(t)=\beta(t)dB(t) + \int_{ \mathbb{R^*} } \kappa(t,\zeta) \widetilde{N}(dt,d\zeta), \quad E[F]= F(0),
\end{align*}
for unique $\mathbb{F}$-predictable processes $\beta(t) \in L^2(\lambda_0 \times P),
\kappa(t,\zeta) \in L^2(\lambda_0 \times \nu \times P)$, where $\lambda_0$ denotes Lebesgue measure.\\
Note that $\beta$ and $\kappa$ depend linearly on $F$ (they can be given in terms of Malliavin derivatives of $F$).

Then by the It\^{o} formula for jump diffusions (see e.g. Theorem 1.14 in \citep*{OS}) 
we get
\begin{align*}
    d(X(t)F(t))&= X(t) dF(t) + F(t) dX(t) + d[X,F]_t \nonumber\\
    &= X(t)\Big[\beta(t)dB(t) + \int_{ \mathbb{R^*} } \kappa(t,\zeta)\widetilde{N}(dt,d\zeta)\Big]\nonumber\\
    &+X(t)F(t)\Big[\pi(t)\alpha(t) dt + \pi(t) \sigma(t) dB(t) + \int_{ \mathbb{R^*} } \pi(t)\gamma(t,\zeta)\widetilde{N}(dt,d\zeta)\Big]\nonumber\\
    & + X(t)\Big[\pi(t)\sigma(t)\beta(t) dt + \int_{ \mathbb{R^*} } \pi(t)\gamma(t,\zeta)\kappa(t,\zeta)\widetilde{N}(dt,d\zeta)\nonumber\\
 &+  \int_{ \mathbb{R^*} } \pi(t)\gamma(t,\zeta)\kappa(t,\zeta)\nu(d\zeta)dt\Big]. 
\end{align*}
Hence
\begin{align*}
    E[X(T)F(T)]&=z F(0)+\int_0^T E\Big[ X(t)\Big\{F(t)\pi(t)\alpha(t)
    + \pi(t)\sigma(t)\beta(t) \nonumber \\
    &+ \int_{ \mathbb{R^*} } \pi(t)\gamma(t,\zeta)\kappa(t,\zeta)\nu(d\zeta)\Big\}\Big]dt. 
\end{align*}
Similarly,
\begin{align*}
E[X^2(T)]= z^2 + \int_0^T E\Big[ X^2(t)\Big\{2 \pi(t)\alpha(t) + 
\pi^2(t)\sigma^2(t) + \int_{ \mathbb{R^*} } \pi^2(t) \gamma^2(t,\zeta) \nu(d\zeta)\Big\} \Big]dt,
\end{align*}
and
\begin{align*}
    E[F^2(T)]=F(0)^2+\int_0^TE\Big[\beta^2(t) + \int_{ \mathbb{R^*} }\kappa^2(t,\zeta) \nu(d\zeta)\Big]dt. 
\end{align*}
This gives
\small
\begin{align*}
    J(\pi)&=E\Big[\frac{1}{2} (X(T) -F)^2\Big]= \frac{1}{2}\Big(E[X^2(T)] - 2E[X(T)F(T)]+E[F^2(T)]\Big)\nonumber\\
    &=\frac{1}{2}(z-F(0))^2 + \frac{1}{2}E\Big[\int_0^T\Big\{2 X^2(t)\pi(t)\alpha(t)
    +  X^2(t)\pi^2(t) \Big(\sigma^2(t)+\int_{ \mathbb{R^*} } \gamma^2(t,\zeta) \nu(d\zeta)\Big)\nonumber\\
    &-2X(t)\pi(t)\Big(F(t) \alpha(t)
    + \sigma(t)\beta(t) +\int_{ \mathbb{R^*} }\gamma(t,\zeta)\kappa(t,\zeta) \nu(d\zeta)\Big)\nonumber\\ 
    &+\beta^2(t) + \int_{ \mathbb{R^*} }\kappa^2(t,\zeta) \nu(d\zeta)\Big\} dt\Big].
\end{align*}
We can minimize $J(\pi)$ by minimizing the $dt$-integrand pointwise for each $t$. This gives the following result:
\begin{theorem} \label{th2.1} 
Recall that we have made the assumption \eqref{1.2}.\\
a) For given initial value $X(0)=z > 0$ the portfolio $\widehat{\pi}=\widehat{\pi}_{z}$ which minimises
\begin{align*}
   \pi \mapsto  E\Big[\frac{1}{2}(X_{z,\pi}(T)-F)^2\Big]
\end{align*}
is given in feedback form with respect to $X(t)=X_{z,\widehat{\pi}}$ by
\begin{align}
    \widehat{\pi}(t)&=\widehat{\pi}(t,X(t))= \frac{F(t)\alpha(t) +\sigma(t)\beta(t)+\int_{ \mathbb{R^*} }\gamma(t,\zeta)\kappa(t,\zeta)\nu(d\zeta) - X(t)\alpha(t)}
    {X(t) \Big( \sigma^2(t) + \int_{ \mathbb{R^*} } \gamma^2(t,\zeta) \nu(d\zeta) \Big)}, \label{3.9a}
\end{align}
or, equivalently,
\begin{align*}
    \widehat{\pi}(t)X(t)=G(t)[X(t)-F(t)]+\frac{\sigma(t)\beta(t)+\int_{\mathbb{R^*}}\gamma(t,\zeta)\kappa(t,\zeta)\nu(d\zeta)}{\sigma^2(t)+\int_{\mathbb{R^*}}\gamma^2(t,\zeta)\nu(d\zeta)},
\end{align*}
where
\begin{equation} \label{(6)}
    G(t)=-\alpha(t)\Big(\sigma^2(t)+\int_{\mathbb{R^*}} \gamma^2(t,\zeta) \nu(d\zeta)\Big)^{-1}.
\end{equation}
b) Given an initial value $z>0$, the corresponding optimal wealth $X_{\widehat{\pi}}(t)=\widehat{X}(t)$ solves the SDE 
\begin{align} \label{2.10}
    d\widehat{X}(t)&=\widehat{X}(t)\widehat{\pi}(t,\widehat{X}(t))\Big[\alpha(t) dt + \sigma(t) dB(t) + \int_{ \mathbb{R^*} }\gamma(t,\zeta)\widetilde{N}(dt,d\zeta)\Big]\nonumber\\
   &=\frac{F(t)\alpha(t) +\sigma(t)\beta(t)+\int_{ \mathbb{R^*} }\gamma(t,\zeta)\kappa(t,\zeta)\nu(d\zeta) - X(t)\alpha(t)}
    {\sigma^2(t) + \int_{ \mathbb{R^*} } \gamma^2(t,\zeta) \nu(d\zeta)}\times\nonumber\\
    &\times \Big[\alpha(t) dt + \sigma(t) dB(t)+\int_{ \mathbb{R^*} }\gamma(t,\zeta)\widetilde{N}(dt,d\zeta)\Big].
\end{align}
\end{theorem}
\subsection{The optimal initial endowment}
Completing the Stackelberg game, we now proceed to find the initial endowment $\widehat{z}$ which leads to a response $\widehat{\pi}=\pi_{\widehat{z}}$ that is optimal for Problem \ref{mv}, in a sense that  $J(\widehat{z},\widehat{\pi}) \leq J(z,\pi)$ over all pairs $(z,\pi)$. We shall first find the explicit solution for $X$.

 Writing $X=\widehat{X}$ for notational simplicity, equation \eqref{2.10} is of the form
\begin{equation}\label{(1)}
dX\left( t\right) = C\left( t\right) d\Lambda _{t}+X\left( t\right)
d\Gamma _{t} ;\quad
X\left( 0\right) = z,
\end{equation}%
where%
\begin{equation}
C\left( t\right) =\frac{F\left( t\right) \alpha \left( t\right) +\sigma
\left( t\right) \beta \left( t\right) +\int_{ \mathbb{R^*} }\gamma \left(
t,\zeta\right) \kappa\left( t,\zeta\right) \nu \left( d\zeta\right) }{ \sigma
^{2}\left( t\right) +\int_{ \mathbb{R^*} }\gamma ^{2}\left( t,\zeta\right) \nu
\left( d\zeta\right) },  \label{(2)}
\end{equation}%
\begin{equation}
d\Lambda _{t}=\alpha \left( t\right) dt+\sigma \left( t\right) dB\left(
t\right) +\int_{ \mathbb{R^*} }\gamma \left( t,\zeta\right) \widetilde{N}\left(
dt,d\zeta\right),  \label{(3)}
\end{equation}%
\begin{equation*}
d\Gamma _{t}=\alpha _{1}\left( t\right) dt+\sigma _{1}\left( t\right)
dB\left( t\right) +\int_{ \mathbb{R^*} }\gamma _{1}\left( t,\zeta\right) \widetilde{N%
}\left( dt,d\zeta\right) ,  
\end{equation*}%
with%
\begin{equation*}
\alpha _{1}\left( t\right) =G\left( t\right) \alpha \left( t\right); \quad \sigma
_{1}\left( t\right) =G\left( t\right) \sigma \left( t\right); \quad \gamma
_{1}\left( t,\zeta\right) =G\left( t\right) \gamma \left( t,\zeta\right).
\end{equation*}%
We rewrite (\ref{(1)}) as%
\begin{equation}
dX\left( t\right) -X\left( t\right) d\Gamma _{t}=C\left( t\right) d\Lambda
_{t}, \label{(7)}
\end{equation}%
and multiply this equation by a process of the form%
\begin{equation*}
Y_{t}=Y_{t}^{\left( \rho ,\lambda ,\theta \right) }=\exp \left(
A_{t}^{\left( \rho ,\lambda ,\theta \right) }\right),  
\end{equation*}%
with%
\begin{equation*}
A_{t}^{\left( \rho ,\lambda ,\theta \right) }=\int_{0}^{t}\rho \left(
s\right) ds+\int_{0}^{t}\lambda \left( s\right) dB\left( s\right)
+\int_{0}^{t}\int_{ \mathbb{R^*} }\theta \left( s,\zeta\right) \widetilde{N}\left(
ds,d\zeta\right),
\end{equation*}%
where $\rho ,\lambda $ and $\theta $ are processes to be determined.\\
Then (\ref{(7)}) gets the form%
\begin{equation}
Y_{t}dX\left( t\right) -Y_{t}X\left( t\right) d\Gamma _{t}=Y_{t}C\left(
t\right) d\Lambda _{t}.  \label{(10)}
\end{equation}%
We want to choose $\rho ,\lambda $ and $\theta $ such that $Y_t$ becomes an integrating factor, in the sense that
\begin{equation*}
d\left( Y_{t}X\left( t\right) \right) =Y_{t}dX\left( t\right) -Y_{t}X\left(
t\right) d\Gamma _{t}+\text{terms not depending on }X. 
\end{equation*}%
To this end, note that by the It\^{o} formula for L\' evy processes, we have
\begin{eqnarray*}
dY_{t} &=&Y_{t}\left[ \rho \left( t\right) dt+\lambda \left( t\right)
dB\left( t\right) \right] +\frac{1}{2}Y_{t}\lambda ^{2}\left( t\right) dt
 \\
&&+\int_{ \mathbb{R^*} }\left\{ \exp \left( A_{t}+\theta \left( t,\zeta\right)
\right) -\exp \left( A_{t}\right) -\exp \left( A_{t}\right) \theta \left(
t,\zeta\right) \right\} \nu \left( d\zeta\right) dt  \nonumber \\
&&+\int_{ \mathbb{R^*} }\left\{ \exp \left( A_{t}+\theta \left( t,\zeta\right)
\right) -\exp \left( A_{t}\right) \right\} \widetilde{N}\left( dt,d\zeta\right) 
\nonumber \\
&=&Y_t \left[ \left\{ \rho \left( t\right) +\frac{1}{2}\lambda ^{2}\left(
t\right) +\int_{ \mathbb{R^*} }\left( e^{\theta \left( t,\zeta\right) }-1-\theta
\left( t,\zeta\right) \right) \nu \left( d\zeta\right) \right\} dt\right.   \nonumber
\\
&&\left. +\lambda \left( t\right) dB\left( t\right) +\int_{ \mathbb{R^*} }\left(
e^{\theta \left( t,\zeta\right) }-1\right) \widetilde{N}\left( dt,d\zeta\right)
\right] .  \nonumber
\end{eqnarray*}%
Therefore, again by the It\^{o} formula, using \eqref{(1)},
\begin{eqnarray*}
d\left( Y_{t}X\left( t\right) \right)  &=&Y_{t}dX\left( t\right) +X\left(
t\right) dY_{t}+d\left[ X,Y\right] _{t}   \\
&=&Y_{t}dX\left( t\right) +Y_{t}X\left( t\right) \left[ \left\{ \rho +\frac{1%
}{2}\lambda ^{2}+\int_{ \mathbb{R^*} }\left( e^{\theta \left( t,\zeta\right)
}-1-\theta \left( t,\zeta\right) \right) \nu \left( d\zeta\right) \right\} dt\right. 
\nonumber \\
&&\left. +\lambda \left( t\right) dB\left( t\right) +\int_{ \mathbb{R^*} }\left(
e^{\theta \left( t,\zeta\right) }-1\right) \widetilde{N}\left( dt,d\zeta\right) %
\right]   \nonumber \\
&&+Y_{t}X\left( t\right) \left[ \left\{ \lambda \left( t\right) \sigma
_{1}\left( t\right) +\int_{ \mathbb{R^*} }\left( e^{\theta \left( t,\zeta\right)
}-1\right) \gamma _{1}\left( t,\zeta\right) \nu \left( d\zeta\right) \right\}
dt\right.   \nonumber \\
&&\left. +\int_{ \mathbb{R^*} }\left( e^{\theta \left( t,\zeta\right) }-1\right)
\gamma _{1}\left( t,\zeta\right) \widetilde{N}\left( dt,d\zeta\right) \right]
+Y_{t}C\left( t\right) dK_{t},  \nonumber
\end{eqnarray*}%
where%
\begin{equation}
dK_{t}=\left\{ \lambda \left( t\right) \sigma \left( t\right) +\int_{ \mathbb{R^*} }\left( e^{\theta \left( t,\zeta\right) }-1\right) \gamma \left( t,\zeta\right)
\nu \left( d\zeta\right) \right\} dt+\int_{ \mathbb{R^*} }\left( e^{\theta \left(
t,\zeta\right) }-1\right) \gamma \left( t,\zeta\right) \widetilde{N}\left(
dt,d\zeta\right).  \label{(14)}
\end{equation}%
This gives 
\begin{align*}
&d\left( Y_{t}X\left( t\right) \right) -Y_{t}dX\left( t\right)
+Y_{t}X\left( t\right) d\Gamma _{t}  \\
&=Y_{t}X\left( t\right) \Big[ \Big\{ \rho +\alpha _{1}+\frac{1}{2}\lambda
^{2}+\lambda \sigma_1+\int_{ \mathbb{R^*} }\left( e^{\theta \left( t,\zeta\right) }-1-\theta \left(
t,\zeta\right) \right) \nu \left( d\zeta\right) \Big\}dt   \nonumber \\
&+ \left( \lambda \left( t\right) +\sigma _{1}\left( t\right) \right)
dB\left( t\right) +\int_{ \mathbb{R^*} }\left( e^{\theta \left( t,\zeta\right)
}-1\right) \gamma _{1}\left( t,\zeta\right) \nu \left( d\zeta\right) dt 
\nonumber \\
& +\int_{ \mathbb{R^*} }\Big\{ \left( e^{\theta \left( t,\zeta\right)
}-1\right) \left( 1+\gamma _{1}\left( t,\zeta\right) \right) +\gamma _{1}\left(
t,\zeta\right) \Big\} \widetilde{N}\left( dt,d\zeta\right)   \nonumber \\
&+Y_{t}C\left( t\right) 
dK_t. \nonumber
\end{align*}%
Choose $\theta \left( t,\zeta\right) =\widehat{\theta }\left( t,\zeta\right) $, such
that 
\[
( e^{\theta \left( t,\zeta\right) }-1) \left( 1+\gamma _{1}\left(
t,\zeta\right) \right) +\gamma _{1}\left( t,\zeta\right) =0,
\]%
i.e.
\begin{equation*}
\widehat{\theta }\left( t,\zeta\right) =-\ln \left( 1+\gamma _{1}\left(
t,\zeta\right) \right).  
\end{equation*}%
Next, choose $\lambda \left( t\right) =\widehat{\lambda }\left( t\right) $
such that 
\begin{equation*}
\widehat{\lambda }\left( t\right) =-\sigma _{1}\left( t\right) .
\end{equation*}%
Finally, choose $\rho \left( t\right) =\widehat{\rho }\left( t\right) $, such
that 
\begin{align*}
\widehat{\rho }( t) &=-\Big[ \alpha _{1}( t) +\frac{1}{%
2}\sigma _{1}^{2}( t) -\sigma_1^2(t)+\int_{ \mathbb{R^*} }\Big( e^{\widehat{\theta 
}( t,\zeta) }-1-\widehat{\theta }( t,\zeta) +( e^{%
\widehat{\theta }( t,\zeta) }-1) \gamma _{1}( t,\zeta)
\Big) \nu  d\zeta) \Big]\nonumber\\
&=-\left[ \alpha _{1}\left( t\right) -\frac{1}{%
2}\sigma _{1}^{2}\left( t\right) +\int_{ \mathbb{R^*} }\Big( \ln(1+\gamma_1(t,\zeta)) - \gamma_1(t,\zeta)
\Big) \nu ( d\zeta) \right].
\end{align*}%
Then 
\begin{align}
  \widehat{A}_t:&=A_t^{(\widehat{\rho},\widehat{\lambda},\widehat{\theta})}\nonumber\\
  &=-\Big[\int_0^t\{ \alpha_1(s)-\frac{1}{2}  \sigma_1^2(s) + \int_{ \mathbb{R^*} }(\ln(1+\gamma_1(s,\zeta))-\gamma_1(s,\zeta))\nu(d\zeta)\} ds \nonumber\\
  &+\int_0^t \sigma_1(s)dB(s)+\int_0^t \int_{ \mathbb{R^*} } \ln(1+\gamma_1(s,\zeta))\widetilde{N}(ds,d\zeta) \Big],\label{2.20a}
\end{align}
with $\widehat{Y}_{t}=Y_{t}^{\left( \widehat{\rho },\widehat{\lambda },%
\widehat{\theta }\right)} = \exp(\widehat{A}_t)$, we have, by (\ref{(14)})%
\begin{eqnarray*}
d\left( \widehat{Y}_{t}X\left( t\right) \right) -\widehat{Y}_{t}dX\left(
t\right) +\widehat{Y}_{t}X\left( t\right) d\Gamma _{t}  
=\widehat{Y}_{t}C\left( t\right) dK_{t}.  
\end{eqnarray*}%
Substituting this into (\ref{(10)}), we get 
\[
d\left( \widehat{Y}_{t}X\left( t\right) \right) -\widehat{Y}_{t}C\left(
t\right) dK_{t}=\widehat{Y}_{t}C\left( t\right) d\Lambda _{t},
\]%
which we integrate to, since $\widehat{Y}_0=1$, 
\[
\widehat{Y}_{t}X\left( t\right) =z+\int_{0}^{t}\widehat{Y}_{s}C\left(
s\right) d\left( K_{s}+\Lambda _{s}\right).
\]%
Solving for $X\left( t\right) $, we obtain the following:
\begin{theorem}\label{th2.3}
Given initial value $z$ the corresponding optimal wealth process $\widehat{X}_z(t)$ is given by
\begin{align}
\widehat{X}_z(t) &=z\widehat{Y}_{t}^{-1}+\widehat{Y}_{t}^{-1}\int_{0}^{t}
\widehat{Y}_{s}C( s) d( K_{s}+\Lambda
_{s})    \nonumber\\
&=z\exp ( -\widehat{A}_t) +\exp ( -\widehat{A}_t)\int_{0}^{t}\exp (\widehat{A}_s)
C\left( s\right) d\left( K_{s}+\Lambda _{s}\right).   \label{(20)}\end{align}

\end{theorem}
In particular, note that%
\begin{equation}
\frac{d}{dz}\widehat{X}_{z}\left( t\right) =\exp ( -\widehat{A}_{t}) . \label{(21)}
\end{equation}

Going back to our problem, choose $z \in \mathbb{R}$ and let $\widehat{\pi}_z$ be the corresponding optimal hedging portfolio given by \eqref{3.9a} and let $\widehat{X}_z$ be the corresponding optimal wealth process given by \eqref{2.10} and \eqref{(20)} respectively.  Then
\begin{align*}
    \inf_{z,\pi}J(z,\pi)=\inf_{z,\pi}E\Big[\frac{1}{2}(X_{z,\pi}(T)-F)^2\Big]=\inf_{z}E\Big[\frac{1}{2}(X_{z,\widehat{\pi}_z}(T) -F)^2\Big]=\inf_{z}E\Big[\frac{1}{2}(\widehat{X}_{z}(T) -F)^2\Big].
\end{align*}

\noindent Note that, if we define
\begin{align*}
    R_t&:=\exp(-\widehat{A}_t)=\exp\Big[\int_0^t\Big\{ \alpha_1(s)-\frac{1}{2}  \sigma_1^2(s) + \int_{ \mathbb{R^*} }\Big(\ln(1+\gamma_1(s,\zeta))-\gamma_1(s,\zeta)\Big)\nu(d\zeta)\Big\} ds \nonumber\\
  &+\int_0^t \sigma_1(s)dB(s)+\int_0^t \int_{ \mathbb{R^*} } \ln(1+\gamma_1(s,\zeta))\widetilde{N}(ds,d\zeta) \Big],
\end{align*}
and
\small
\begin{align*}
    Z^*(t)&:=\exp\Big(-\int_0^t \alpha_1(s)ds\Big)R_t\\
    &=\exp\Big[\int_0^t\Big\{-\frac{1}{2}  \sigma_1^2(s) + \int_{\mathbb{R^*}}(\ln(1+\gamma_1(s,\zeta))-\gamma_1(s,\zeta))\nu(d\zeta)\Big\} ds \nonumber\\
  &+\int_0^t \sigma_1(s)dB(s)+\int_0^t \int_{ \mathbb{R^*} } \ln(1+\gamma_1(s,\zeta))\widetilde{N}(ds,d\zeta) \Big],
\end{align*}
then we can verify by the It\^{o} formula that
\begin{align*}
    dR_t&=R_t\Big(\alpha_1(t)dt + \sigma_1(t) dB(t) + \int_{ \mathbb{R^*} } \gamma_1(t,\zeta) \widetilde{N}(dt,d\zeta)\Big)\nonumber\\
    &=R_t G(t)\Big(\alpha(t)dt + \sigma(t) dB(t) + \int_{ \mathbb{R^*} } \gamma(t,\zeta) \widetilde{N}(dt,d\zeta)\Big)\nonumber\\
    &= R_t G(t) S^{-1}(t)dS(t),
    \end{align*}
    and
 \begin{align} \label{Z*} 
 dZ^{*}(t)&=Z^{*}(t) G(t)\Big( \sigma(t) dB(t) + \int_{ \mathbb{R^*} } \gamma(t,\zeta) \widetilde{N}(dt,d\zeta)\Big).
\end{align}
\begin{proposition}
$Z^{*}(\cdot)$ is a $P$-martingale. (See \eqref{KS}.) In particular, it follows that $Q^{*}$ defined by
\begin{align}\label{Q*}
    dQ^{*}(\omega)=Z^{*}(T) dP(\omega) \text { on } \mathcal{F}_T.
\end{align}
is an EMM for $S(\cdot)$.
\end{proposition}
\dproof To see this, we verify that the coefficients 
$\theta_0(t):=G(t)\sigma(t)$ and $\theta_1(t,\zeta):=G(t)\gamma(t,\zeta)$ satisfy condition \eqref{emm}:

 \begin{align*}
&\alpha(t)+\theta_{0}(t)\sigma(t)+\int_{
\mathbb{R^*}
}\theta_{1}(t,\zeta)\gamma(t,\zeta)\nu(d\zeta)\nonumber\\
&=\alpha(t)-\frac{\alpha(t)}{\sigma^2(t) + \int_{ \mathbb{R^*} } \gamma^2(t,\zeta) \nu(d\zeta)}\sigma^2(t) -\frac{\alpha(t)}{\sigma^2(t) + \int_{ \mathbb{R^*} }\gamma^2(t,\zeta) \nu(d\zeta)}\int_{ \mathbb{R^*} }\gamma^2(t,\zeta) \nu(d\zeta)\nonumber\\
&= \alpha(t) -\alpha(t)=0\text{, }\quad t\geq 0.
\end{align*}   
\fproof

Using this, we obtain the following, which is the main result in this section:
\begin{theorem}\label{th2.5}
(i) The unique minimal variance price $\widehat{z}$ of a European option with terminal payoff $F$ at time $T$ is given by
\begin{align} \label{3.14a}
\widehat{z}= \frac{E\Big[ exp(-\widehat{A}_T)\Big(F -exp(-\widehat{A}_T) \int_0^T exp(\widehat{A}_s)C\left( s\right) d\left( K_{s}+\Lambda _{s}\right)\Big)\Big]}{E[exp(-2\widehat{A}_T)]},
\end{align}
where $\widehat{A}_T$ is given by \eqref{2.20a}, $C(s), \Lambda_s$ are given by \eqref{(2)}, \eqref{(3)} respectively, and $K$ is given by \eqref{(14)}.\\
(ii) Assume that the coefficient $\alpha(t)$, $\sigma(t)$ and $\gamma(t,\zeta)$ are bounded and deterministic functions.
Then
\begin{align}
 \widehat{z}=E_{Q^{*}}[F],  \label{2.43} 
\end{align}
where $Q^{*}$ is the EMM measure given by \eqref{Q*}.
\end{theorem}

\dproof 
(i) To minimize $J_0(z):= E\Big[\frac{1}{2} (\widehat{X}_{z}(T)-F)^2\Big]$ with respect to $z$ we get by \eqref{(20)} and \eqref{(21)} that
\begin{align}
\frac{d}{dz} J_0(z)
&=E\Big[(\widehat{X}_{z}(T) -F)\frac{d}{dz}\widehat{X}_{z}(T)\Big] \nonumber\\
&=E[(\widehat{X}_z(T) -F)exp(-\widehat{A}_T)] \label{2.43a}\\
&=E\Big[\Big(zexp(-\widehat{A}_T) +exp(-\widehat{A}_T)\int_0^T exp(\widehat{A}_s)C\left( s\right) d\left( K_{s}+\Lambda _{s}\right)-F\Big)exp(-\widehat{A}_T)\Big].\nonumber
\end{align}
This is $0$ if and only if \eqref{3.14a} holds.
\vskip 0.2cm
\noindent (ii) 
By \eqref{2.43a}, we get
\begin{align*}
 &E[\widehat{X}_z(T)\exp(-\widehat{A}_T)]=E[F \exp(-\widehat{A}_T)],\nonumber\\
 \text{ i.e. }\nonumber\\
 &E\Big[\widehat{X}_z(T)\exp\Big(\int_0^T \alpha_1(s)ds\Big) Z^{*}(T)\Big]=E\Big[F\exp\Big(\int_0^T \alpha_1(s)ds\Big) Z^{*}(T)\Big].
\end{align*}
If $\alpha_1$ is deterministic, we cancel out the factor $\exp\Big(\int_0^T \alpha_1(s)ds\Big)$ and \eqref{2.43} follows.
\fproof

\begin{remark}
In particular, if $F$ is a deterministic constant. Then $F(t)=F=E[F]$ for all $t$, and $\beta=\kappa =0$. Hence the optimal portfolio is given in feedback form by
$$\widehat{\pi}(t)X(t)=G(t)[X(t)-F].$$
Assume, for example, that $\alpha(t) > 0$. Then $G(t) < 0$, and we see that if $X(t) < F$, then $\widehat{\pi}(t)X(t) >0$, and hence the optimal portfolio pushes $X(t)$ upwards towards F. Similarly, if $X(t) > F$ then $\widehat{\pi}(t)X(t) < 0$ and the optimal push of $X(t)$  is downwards towards $F$. This is to be expected since the portfolio tries to minimize the terminal variance $E[(X(T)-F)^2]$.
\vskip 0.2cm
Moreover, if we start at $z=X(0)=F$, we can choose $\pi=0$ and this gives 
$J(z,\pi)=J(F,0)=E[\frac{1}{2}(X(T)-F)^2]=E[\frac{1}{2}(F-F)^2]=0$, which is clearly optimal. By uniqueness of $(\widehat{z},\widehat{\pi})$ we conclude that $(\widehat{z},\widehat{\pi})=(F,0)$ is the optimal pair in this case.  
\end{remark}

\begin{remark}
The price $\widehat{z}$ is an arbitrage-free price of $F$,
follows by Theorem \ref{th2.5} (ii). 
\end{remark}
\begin{remark} 
Note that, as remarked earlier, the coefficients $\beta$ and $\kappa$ depend linearly on $F$. Therefore it follows from the formula \eqref{3.14a} that the map $\Phi: L^2(P,\mathcal{F}_T) \mapsto \mathbb{R}$ defined by
$$\Phi(F)= \widehat{z}; \quad F \in L^2(P,\mathcal{F}_T)$$
is linear and bounded. By the Riesz representation theorem this map can be represented by a random variable $Z \in L^2(P,\mathcal{F}_T)$, in the sense that
\begin{align*}
\widehat{z}=E[ ZF]; \quad F \in L^2(P,\mathcal{F}_T).
\end{align*}
Therefore, if we define the (signed) measure $\tilde{Q}$ on $\mathcal{F}_T$ by
\begin{align*}
d\tilde{Q}=Z dP
\end{align*}
then 
\begin{align*}
\widehat{z}=E_{\tilde{Q}}[F].
\end{align*}
\end{remark}
Comparing with the Schweizer variance-optimal pricing measure $\tilde{P}$ \eqref{sw} we conclude the following:
\begin{corollary}
\begin{myenumerate}
\item
$\tilde{P}=\tilde{Q}$\\
In particular, $\tilde{P}$ always exists in this market, without any non-arbitrage conditions.
\item
Moreover, we have\\
$\tilde{P}=Q^{*}$,
which is a positive EMM.
\end{myenumerate}
\end{corollary}

\subsection{Example: European call option}
We give some details about how to compute the minimal variance price $\widehat{z}$ explicitly in the case of a European call option that will used in Section \ref{sec:dl}.
\begin{myenumerate}
    \item 
Note that the term $C(s)$ in Theorem \ref{th2.5} depends on the coefficients $\beta$ and $\kappa$ in the It\^o representation of $F$. These coefficients can for example be found by using the generalised Clark-Ocone formula for L\' evy processes, extended to $L^2(P)$. See Theorem 12.26 in \citep*{DOP}. \\
Let us find these coefficients in the case of a European call option, where
$$F=(S(T)-K)^{+},$$
where $K$ is a given exercise price. In this case $F(\omega)$ represents the payoff at time $T$ (fixed) of a (European call) option which gives the owner the right to buy the stock with value $S(T,\omega)$ at a fixed exercise price $K$. Thus if $S(T,\omega)>K$ the owner of the option gets the profit $S(T,\omega)-K$ and if $S(T,\omega)\leq K$ the owner does not exercise the option and the profit is 0. Hence in this case
$$F(\omega)=(S(T,\omega)-K)^{+}.$$
Thus, we may write
$$F(\omega)=f(S(T,\omega)),$$
where

$$f(x)=(x-K)^{+}.$$

The function $f$ is not differentiable at $x=K$, so we cannot use the chain rule directly to evaluate $D_tF$. However, we can approximate f by $C^1$ functions $f_n$ with the property that 
$$f_n(x)=f(x) \quad \text{ for } \quad |x-K| \geq \frac{1}{n},$$
and 
$$0\leq f'_n(x)\leq 1 \text{ for all } x.$$
Putting
$$F_n(\omega)=f_n(S(T,\omega)),$$
we see
$$D_tF(\omega)=\lim_{n \to +\infty} D_tF_n(\omega).$$

We get
\begin{align*}
    \beta(t)= E[D_t F | \mathcal{F}_t], 
    \kappa(t,\zeta)=E[D_{t,\zeta}F | \mathcal{F}_t],
\end{align*}
where $D_t F$ and $D_{t,\zeta} F$ denote the generalised Malliavin derivatives (also called the Hida-Malliavin derivative) of $F$ at  $t$ and $(t,\zeta)$ respectively, with respect to $B(\cdot)$ and $N(\cdot,\cdot)$, respectively. Combining this with the chain rule for the Hida-Malliavin derivative and the Markov property of the process $S(\cdot)$, and assuming for simplicity that $\sigma$ is constant and $\gamma(t,\zeta)=\gamma(\zeta)$ does not depend on $t$,  we obtain the following for $\beta$:
\begin{align}\label{beta}
    \beta(t)&= E^{S_0}\Big[\mathbbm{1}_{[K,\infty)}(S(T))\sigma S(T) \Big|\mathcal{F}_t\Big]\nonumber\\
    &=E^{S(t)}\Big[\mathbbm{1}_{[K,\infty)}(S(T-t))\sigma S(T-t)\Big].
\end{align}

To find the corresponding result for $\kappa$ we first use the chain rule for $D_{t,\zeta}$ and get
\begin{align*}
    D_{t,\zeta}S(T)&=D_{t,\zeta}\Big[S_0 \exp \Big(\alpha T -\frac{1}{2}\sigma^2 T+ \sigma B(T) +\int_{\mathbb{R^*}}(\log(1+\gamma(\zeta))-\gamma(\zeta))\nu(d\zeta)T\nonumber\\&+\int_0^t\int_{\mathbb{R^*}}\ln(1+\gamma(\zeta))\widetilde{N}(ds,d\zeta) \Big)\Big]=S(T)\gamma(\zeta).
\end{align*}
Then we obtain
\begin{align} \label{kappa}  \kappa(t,\zeta)&=E^{S_0}\Big[\mathbbm{1}_{[K,\infty)}(S(T)+D_{t,\zeta}S(T))-\mathbbm{1}_{[K,\infty)}(S(T)) \Big|\mathcal{F}_t\Big]\nonumber\\
    &=E^{S_0}\Big[\mathbbm{1}_{[K,\infty)}(S(T)+\gamma(\zeta)S(T))
    -\mathbbm{1}_{[K,\infty)}(S(T)) \Big|\mathcal{F}_t\Big] \nonumber\\&=E^{S(t)}\Big[\mathbbm{1}_{[K,\infty)}(S(T-t)+\gamma(\zeta)S(T-t)) - \mathbbm{1}_{[K,\infty)}(S(T-t))\Big],
\end{align}
where in general $E^{y}[h(S(u))]$ means $E[h(S^{y}(u))]$, i.e.   expectation when $S$ starts at $y$.
\item Since the coefficients $\alpha, \sigma$ and $\gamma$ of the process $S$ are deterministic and bounded, we compute numerically the minimal variance price
$$\widehat{z}=E\Big[(S(T)-K)^{+}Z_T^{*}\Big]$$
of a European call option with payoff  
\begin{align*}
    F=(S(T)-K)^{+}= E[F]+\int_0^T \beta(t)dB(t)+\int_0^T \int_{\mathbb{R^*}} \kappa(t,\zeta) \widetilde{N}(dt,d\zeta),
\end{align*}
where $\beta,\kappa$ are given by \eqref{beta},\eqref{kappa}, respectively, we use the It\^o formula combined with \eqref{Z*} to obtain
\begin{align*}
    \widehat{z}=E[F Z_T^{*}]=E[F]+ \int_0^T G(t)\Big\{\sigma(t) E[Z_t^{*}\beta(t)] + \int_{\mathbb{R^*}}\gamma(t,\zeta) E[\kappa(t,\zeta) Z_t^{*}] \nu(d\zeta)\Big\} dt,
\end{align*}
where $G(t)$ is given by \eqref{(6)}.
\end{myenumerate}

\subsubsection{Application to Black Scholes market with two independent Brownian motions}\label{sec:BS2B}

In this example, we consider a market driven by two independent Brownian motions, $B_1(t), B_2(t)$: 
\begin{align*}
dS(t)=S(t)[\alpha_0dt +\beta_1dB_1(t) + \beta_2)dB_2(t)],\quad 
S(0) > 0,
\end{align*}
where the coefficients $\alpha_0,\beta_1,\beta_2$ are assumed to be bounded constants.
For a self-financing portfolio $\pi$ and an initial wealth $z$, we have a wealth dynamic
\begin{align*}
dX(t)=X(t)\pi(t)[\alpha_0dt +\beta_1dB_1(t) + \beta_2dB_2(t)],\quad
X(0)=z.
\end{align*}
We want to find the pair $(\widehat{z},\widehat{\pi})$  which minimizes the expected squared hedging error
\begin{align}\label{eq:2bm-loss}
   E\Big[\frac{1}{2}(X(T)-F)^2\Big].
\end{align}
\begin{corollary}
From Theorem \ref{th2.5} (ii), we conclude that
\begin{align}\label{eq:bs2-price}
    E_{Q^{*}}[F] =  E[FZ^*(T)],
\end{align}
where 
\begin{align*}
    Z^*(t) =& \exp \Big(-\frac{1}{2} G^2( \beta_1^2+ \beta_2^2)t
    + G\beta_1 B_1(t)+ G \beta_2B_2(t)\Big),
\end{align*}
and 
\begin{align}\label{eq:G2bm}
    G = -\frac{\alpha_0}{\beta_1^2 + \beta_2^2}.
\end{align}
\end{corollary}

\added{Looking at the above corollary, we can see that process $Z^*$ coincides with the one in the Black-Scholes setting, where the volatility coefficient of the single Brownian motion $B(t)$ is given by $\beta = \sqrt{\beta_1^2 + \beta_2^2}$. Since processes $\beta B(t)$ and $\beta_1 B_1(t) + \beta_2 B_2(t)$ have the same distribution, so do the option payoffs. This yields that the minimal-variance European call option price in an incomplete market with two independent Brownian motions $E_{Q^{*}}[F]$ coincides with the unique Black-Scholes price in a single Brownian motion case for volatility $\beta = \sqrt{\beta_1^2 + \beta_2^2}$.}

\added{Furthermore, we know that the hedging portfolio in the BS model is only a function of time and the underlying asset. Since for the choice of volatility parameter as above, the stocks in both BS and two-Brownian motion model follow the same distribution, we can deduce that for initial endowment $E_{Q^{*}}[F]$ from Equation \eqref{eq:bs2-price} and for the BS hedging portfolio, we get}
\begin{align*}
    \added{E\Big[\frac{1}{2}(X(T)-F)^2\Big] = 0.} 
\end{align*}

\added{The above can, of course, be extended to the arbitrary number of Brownian motions.}

\subsection{Application to Merton model}\label{sec:merton}

Here we consider the special case of the Merton model first proposed in \citep*{M}. It deals with European options in the market modeled by jump diffusion. More precisely, the market consists of 
\begin{description}
\item[(i)] a risk free asset, with unit price $S_{0}(t)=1$ for all $t$,
\item[(ii)] a risky asset, with price $S(t)$ given by
\begin{align}\label{eq:s merton}
    dS(t)=S(t)\Big[\alpha_0 dt + \sigma_0 dB(t) +(y-1)\Tilde{N}(dt,dy)
    \Big], \quad S(0)=S_0 > 0.
\end{align}
\end{description}
Here $\Tilde{N}$ is a compensated Poisson random measure corresponding to the compound Poisson process with intensity $\lambda$ and jump sizes $y = \exp(Y)$, where $Y \sim \mathcal{N}(\mu, \delta^2),$ independent of $B$ and jump times. Heuristically speaking, $y$ represents the absolute jump size, while $\gamma_0(t,y) = y-1$ equals the relative jump size. The coefficients $\alpha_0$ and $\sigma_0$ are assumed to be bounded and deterministic.\\
If we denote $k = E[y-1] = \exp(\mu + \frac{\delta
^2}{2}) -1$ and observe that $y-1 > -1$ a.s. we can use It\^o formula and obtain the explicit solution
\begin{align*}
S(t)& =S_0 \exp\Big( \{\alpha_0-\frac{\sigma_0^2}{2}-\lambda k \}t+\sigma_0 B(t)+\sum^{N(t)}_{i=1} Y_i\Big).
\end{align*}
Merton then proceeds to argue that the jumps in the asset price are not systemic and, therefore, the risk related to the jumps is diversifiable. In other words, this means that all the properties of the jump component of $S(\cdot)$ under the risk neutral measure are the same as under the natural measure. This argument yields the choice $\theta_0 = \frac{\alpha_0}{\sigma_0}$ and $\theta_1(y) = 0$ which results in EMM ${Q}^M$ corresponding to the Radon-Nikod\'ym derivative
\begin{align*}
    Z^M(t) = \exp \Big( -\frac{\alpha_0^2}{2\sigma_0^2} t - \frac{\alpha_0}{\sigma_0} B(t) \Big)
\end{align*}
and enables Merton to obtain the option price  as
\begin{align}\label{eq:merton}
    E_{{Q}^M}[ F] = \sum_{j\geq 0} \frac{e^{- \lambda(k+1)T} ((\lambda(k+1)T)^j}{j!} BS \Big(0,S_0, \sqrt{\sigma_0^2+\frac{j\delta^2}{T}}, \frac{j\mu + j \delta^2/2}{T}- \lambda k\Big),  
\end{align}
where $BS(t, S(t),\sigma_0,r)$ denotes the price of the European call option under the BS model at time $t$, with stock price $S(t)$, volatility $\sigma$, and interest rate $r$.

This approach results in a non-symmetric loss. Merton argues that if no jump occurs, then the owner of the option collects a small profit. However, in a rare case when a jump does occur, the option holder suffers a significant loss. The nonsymmetric loss distribution can, in some instances, be nondesirable. Apart from that, the assumption of diversifiabilty of the jump risk does not hold in practice; for example, market indexes do experience occasional jumps in price. \\

Using the results from the previous section,  we can obtain the optimal portfolio, optimal wealth process, and the unique minimal variance price given by \eqref{3.9a}, \eqref{(20)} and \eqref{3.14a} respectively. To evaluate these expressions, one has to compute coefficients $\beta(t)$ and $\kappa(t,y)$ given by equations \eqref{beta} and \eqref{kappa}. If we denote $L(t) = S(t) \exp\big(( \alpha - \frac{\sigma^2}{2} - \lambda k)(T-t) \big),$ $ H(t) = \sigma B(t) + \sum_{i=1}^{N(t)} Y_i$ we get

\begin{align*}
    \beta(t) &= E^{S(t)}\Big[\mathbbm{1}_{[K,\infty)}(S(T-t))\sigma S(T-t)\Big] \\
    &= \sigma L(t) E[\mathbbm{1}_{[K L(t)^{-1},\infty)}(e^{H(T-t)})e^{H(T-t)}] \\
    &= \sigma L(t) \sum_{j=0}^{\infty} E[\mathbbm{1}_{[K L(t)^{-1},\infty)}(e^{H(T-t)})e^{H(T-t)} \Big| N(T-t) = j] {P}(N(T-t) = j) \\
    &= \sigma L(t) \sum_{j=0}^{\infty} \exp \Big(j \mu + \frac{j \delta^2 + \sigma^2(T-t)}{2} \Big) \cdot \\
    &\cdot \Phi \Big(\frac{j \mu + j \delta^2 + \sigma^2(T-t) - \log(K) + \log(L(t))}{\sqrt{j \delta^2 + \sigma^2 (T-t)}} \Big) \cdot \frac{(\lambda(T-t))^j \exp(-\lambda(T-t))}{j!},
\end{align*}

\begin{align*}
    \kappa(t,y) &= E^{S(t)}\Big[\mathbbm{1}_{[K,\infty)}(S(T-t)+\gamma(y)S(T-t)) - \mathbbm{1}_{[K,\infty)}(S(T-t))\Big] \\
    &= E^{S(t)}\Big[\mathbbm{1}_{[K,\infty)}(S(T-t)+(y-1)S(T-t)) - \mathbbm{1}_{[K,\infty)}(S(T-t))\Big] \\
    &= E^{S(t)}\Big[\mathbbm{1}_{[K,\infty)}(yS(T-t)) - \mathbbm{1}_{[K,\infty)}(S(T-t))\Big] \\
    &= \mathbb{P}\Big(e^{H(T-t)} > \frac{K}{y L(t)} \Big) - \mathbb{P}\Big(e^{H(T-t)} > \frac{K}{L(t)} \Big) \\
    &= \sum_{j=0}^{\infty} \Bigg( \Phi \Big(\frac{\log(K) - \log( L(t)) -j \mu}{\sqrt{j \delta^2 + \sigma^2 (T-t)}} \Big) \\ 
    &- \Phi \Big(\frac{\log(K) - \log(y L(t)) -j \mu}{\sqrt{j \delta^2 + \sigma^2 (T-t)}} \Big) \Bigg) \cdot \frac{(\lambda(T-t))^j \exp(-\lambda(T-t))}{j!},
\end{align*}
where we use the fact that $H(t) | \big\{ N(t) = j \big\} \sim \mathcal{N}\big( j \mu, \sigma^2(T-t) + j\delta^2 \big).$ Unfortunately, this representation is not enough to obtain the desired results explicitly.  
However, we may observe the following. The L\'evy measure associated with the compound process in equation \eqref{eq:s merton} equals $\lambda \mu_y$, where $\mu_y$ is the law of lognormally distributed random variable $y.$ Hence, even though the coefficient $\gamma_0$ is stochastic, we have that $m = \int_{\mathbb{R}^*} \gamma_0(y)^2 \nu(dy)$ is deterministic. Since $\alpha_0$ and $ \sigma_0$ are also deterministic, so is 
\begin{align}\label{eq:G}
    G = - \frac{\alpha_0}{\sigma_0^2 + \lambda m}.
\end{align}
By part (ii) of the Theorem \ref{th2.5}, we have an explicit representation of the EMM $Q^*$ and the corresponding Radon-Nikod\'ym derivative given by 
\begin{align*}
    Z^{*}(t)=\exp \Big(\{-\frac{1}{2}G^2\sigma_0^2 -\lambda G k\}t+G \sigma_0 B(t)+ \sum_{i=1}^{N(t)} \ln(1+G (y_i-1)) \Big).
\end{align*}
Moreover, the minimal variance price $\widehat{z}$ of an option with payoff $F$ at time $T$ is
\begin{align}\label{eq:merton price}
    \widehat{z}=E_{Q^{*}}(F)= E[F Z^{*}(T)]. 
\end{align}

\begin{figure}[H]
\centering
\includegraphics[width=0.5\linewidth]{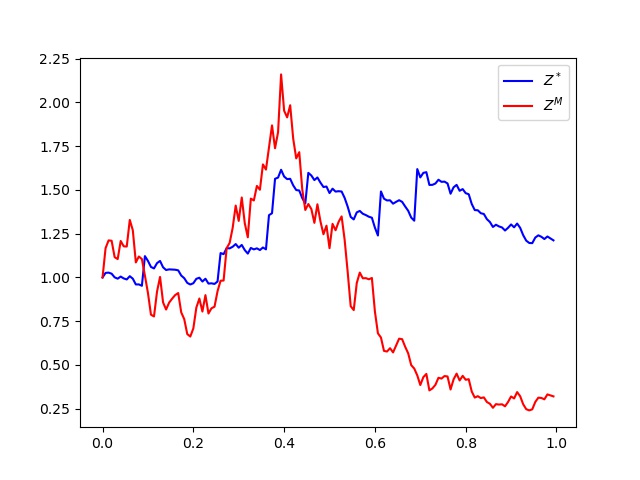}

\caption{One realization of $Z^M$ and $Z^*$ for the choice of parameters specified in Section \ref{sec:num_merton}}
\label{fig:Z}
\end{figure}

\begin{remark}\label{rem:G}
    It must be stated that, in theory, we do not have a guarantee that $G(y_i -1) > -1$ since lognormal distribution takes values on the whole positive line. Consequently, the logarithm may not be defined, in which case we use the alternative formulation 
    \begin{align*}
        Z^*(t) = \Big( \prod_{i=1}^{N(t)} 1+G (y_i-1) \Big) \exp \Big(\{-\frac{1}{2}G^2\sigma_0^2 -\lambda G k\}t+G \sigma_0 B(t) \Big)
    \end{align*}
    from \citep*{lamberton2011introduction}. This results in a signed measure $Q^*$.
    Nonetheless, in practice, for a reasonable choice of parameters, $G(y_i -1) > -1$ holds with probability practically equal to 1, and we do not need to worry about this technical problem.  
\end{remark}

We can not derive the same formula as in \eqref{eq:merton} since the distribution of $\ln(1+G (y-1))$ is unknown; however, we can use the Monte Carlo approach to obtain the option price. This option price obviously differs from Merton's, as does the hedging portfolio that also considers the risk coming from the jumps and results in a symmetric loss. Using the Euler-Maruyama discretization rule, we simulate one realization of $Z^M$ and $Z^*$ and present it in Figure \ref{fig:Z}. We can see that $Z^*$ has upward jumps, which are not present in $Z^M.$ 

\begin{figure}[H]
\centering
\begin{subfigure}{.3\textwidth}
  \centering
  \includegraphics[width=1\linewidth]{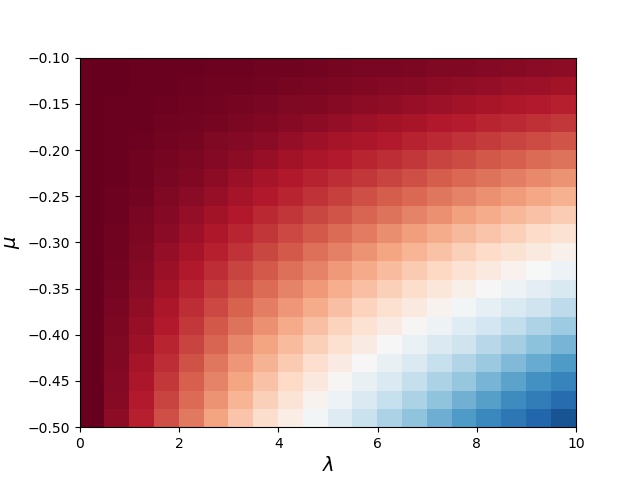}
  \caption{$E_{Q^M}[F]$}
  \label{graph:pmlm}
\end{subfigure}%
\begin{subfigure}{.3\textwidth}
  \centering
  \includegraphics[width=1\linewidth]{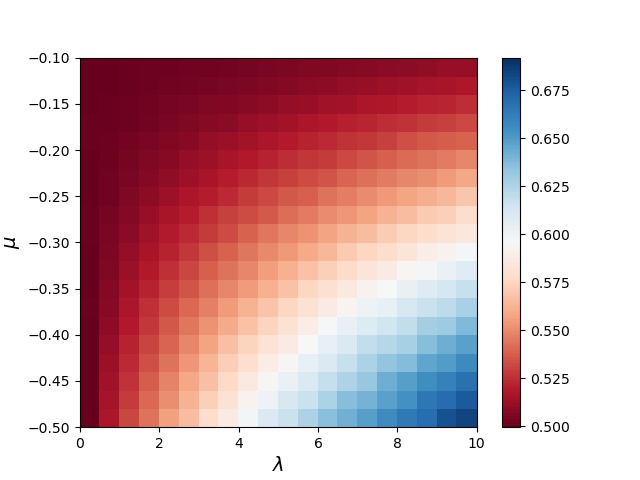}
  \caption{ $E_{Q^*}[F]$}
  \label{graph:polm}
\end{subfigure}
\begin{subfigure}{.3\textwidth}
  \centering
  \includegraphics[width=1\linewidth]{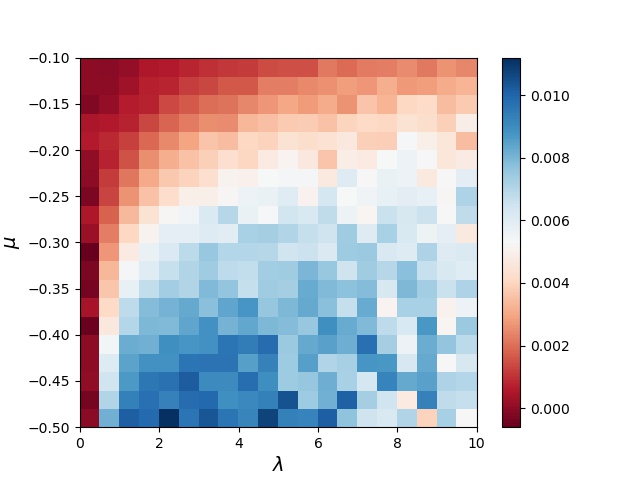}
  \caption{$E_{Q^*}[F] - E_{Q^M}[F]$}
  \label{graph:diflm}
\end{subfigure}
\caption{Option price comparison depending on parameters $\lambda$ and $\mu$, where $S_0 = 1$, $K=0.5,$ $\sigma = 0.2$ and $\delta = 0.05$}
\label{fig:price_lm}
\end{figure}

\begin{figure}[H]
\centering
\begin{subfigure}{.3\textwidth}
  \centering
  \includegraphics[width=1\linewidth]{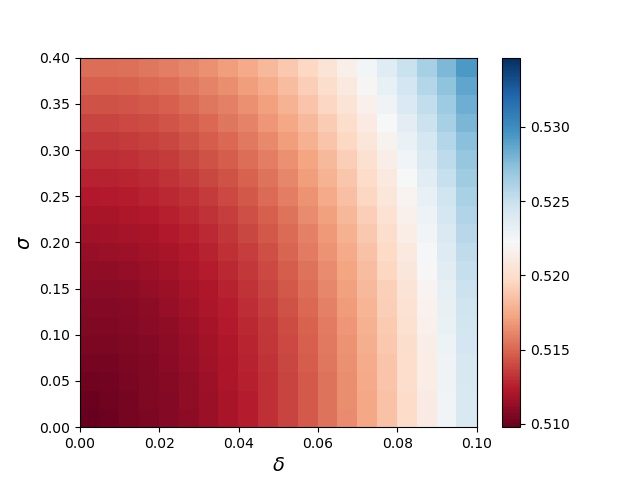}
  \caption{$E_{Q^M}[F]$}
  \label{graph:pmvs}
\end{subfigure}%
\begin{subfigure}{.3\textwidth}
  \centering
  \includegraphics[width=1\linewidth]{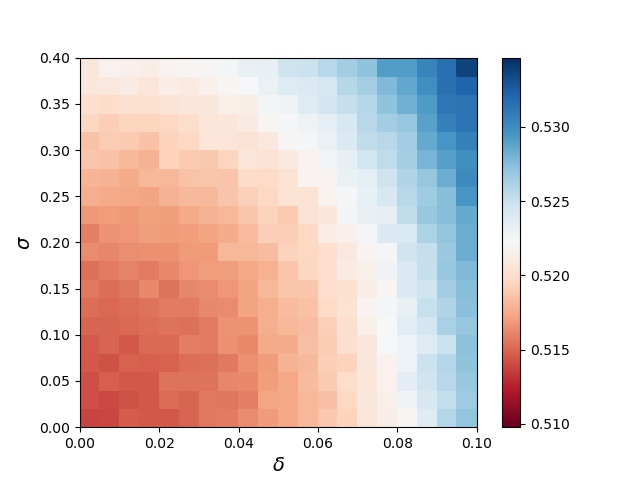}
  \caption{ $E_{Q^*}[F]$}
  \label{graph:povs}
\end{subfigure}
\begin{subfigure}{.3\textwidth}
  \centering
  \includegraphics[width=1\linewidth]{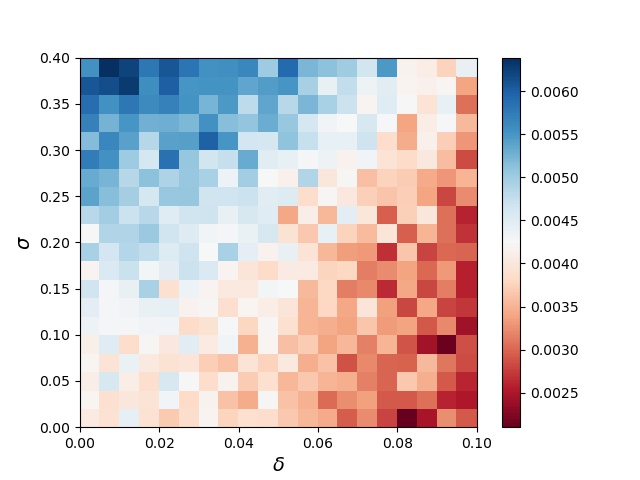}
  \caption{$E_{Q^*}[F] - E_{Q^M}[F]$}
  \label{graph:difvs}
\end{subfigure}
\caption{Option price comparison depending on parameters $\sigma$ and $\delta$, where $S_0 = 1$, $K=0.5,$ $\lambda = 10$ and $\mu = -0.2$}
\label{fig:price_vs}
\end{figure}

In figures \ref{fig:price_lm} and \ref{fig:price_vs}, we can see how the price of the European option changes for both models depending on the parameters $\lambda, \, \mu, \, \sigma, \, \delta.$ The option price in both models grows together with the absolute value of the parameters. Note that option prices coincide with the exclusion of jumps, namely when $\lambda = 0$ since the Merton model devolves into the BS model. It must be noted how the changes in prices are higher in the case described by Figure \ref{fig:price_lm}, where we consider the dependence on parameters $\lambda$ and $\mu$, than in Figure \ref{fig:price_vs}, where $\sigma$ and $\delta$ are taken into consideration, due to those parameters' effect on stock and wealth dynamics. Large values of parameters $\lambda$ and $\mu$ change dynamics directly, while $\sigma$ and especially $\delta$ do it indirectly. In Figure \ref{fig:price_lm}, it is shown how model prices are similar for small absolute values of parameters. When parameters grow, minimal variance price grows faster; however, the difference stabilizes. As a matter of fact, the difference appears to be most significant when $\lambda$ is relatively small, while $|\mu|$ is large. This is because once $\lambda$ is large, the compensated part of the jump component, which is taken into account in both hedging strategies, takes over.  

To continue with observations for the other two parameters, we can see that even though absolute differences are smaller, relatively speaking, models differ more when $\sigma$ and $\delta$ parameters are modified, as shown in Figure \ref{fig:price_vs}. The difference between the prices seems to be biggest when jump sizes are constant while volatility in the model is large.

Numerical methods for obtaining the optimal portfolio are presented in the next section.

\section{Deep Learning Approach}\label{sec:dl}

Here we present the algorithm using deep learning methods that returns both the optimal initial value and the optimal portfolio. In \citep*{FE}, different 
types of neural networks (NN) are compared when studying hedging in discrete-time incomplete markets. It turns out that Long short-term memory (LSTM) networks, which are the particular case of Recurrent neural networks (RNN), outperform standard feed-forward NN. \added{Because the model discussed in the aforementioned paper doesn't incorporate jumps in the price dynamics, we also conducted a comparison between the performance of LSTM and feed-forward neural networks. However, we detected no significant difference between the two approaches. A possible intuitive reason for the good performance of the LSTM network in the jump diffusion models could be its ability to learn the jump dynamics from the previous jump occurrences.} \replaced{Furthermore}{Apart from that}, RNN and, in particular, LSTM networks seem to be the most natural approach when working with time series data and, \replaced{importantly}{furthermore}, allow generalization to non-Markovian models. \added{Hence, we opted to focus exclusively on the LSTM model in the subsequent work.} \replaced{In particular a}{Another} possible extension that would benefit from the use of LSTM networks is if we consider stochastic volatility models. Historical information that LSTM keeps or forgets would help us to store and use knowledge about the behaviour of the volatility parameter. 

\replaced{The performance of our network is to be tested on four models. First, we consider the BS model, where the market is complete. The reason for that is that both option value and hedging portfolio are known and can be compared to our results. Then, we consider three cases of incomplete markets. We start with the BS model with multiple independent Brownian motions and continue with two jump diffusion models, Merton and Kou double exponential models.}{The performance of our network is to be tested on three models. First, we consider the BS model, where the market is complete. The reason for that is that both option value and hedging portfolio are known and can be compared to our results. Then we consider two cases of incomplete markets. We start with the BS model with two independent Brownian motions (BS2B) and continue with the Merton model.}

\subsection{Data Generation}

To generate the data, we discretize time interval $[0,T]$ into $R$ equidistant points $t_i$ at a distance $\Delta_t$. We denote different realizations of the process with $(j)$ in superscript, where $j = [M]$ and $M$ denotes the batch size. When the superscript is omitted, we assume the vector notation. \added{Additionally, we here assume the most general case with $d_B$ dimensional Brownian motion and $d_N$ dimensional Poisson random measure.}

For initial wealth $x_0$ we construct under portfolio $\pi$ the wealth process at time $t_i$ using the update rule 


\begin{equation}\label{eq:discrete}
    x_{i+1} = x_{i} + x_i\pi_i\big[ (\alpha_0-\lambda \added{^\top} k) \Delta_t + \sqrt{\Delta_t}B_i \added{\sigma_0} + \gamma_0 J_i \big], \qquad i \in [R-1].
\end{equation}
\replaced{Here $\alpha_0\in \RB$ and $\sigma_0 \in \RB^{d_B}$ are drift and volatility parameters respectively. Parameter $\gamma_0 \in \{0,1\}$ serves as a dummy variable that indicates whether a model is continuous or not.}{Here $\alpha_0,$ $\sigma_0$ and $\gamma_0$ are drift, volatility, and jump parameters, respectively.} Moreover $B_i$ are $M \added{\times d_B}$ dimensional independent $\mathcal{N}(0,1)$ random variables, while $J_i \added{= \sum_{l=1}^{d_N} J_i^l,}$ \added{where $J_i^l$} are independent and given by 

\begin{equation*}\label{eq:J_i}
    J_{i}\added{^l} = 
\begin{cases}
  e^{Y_i\added{^l}} - 1, & N\added{^l}(t) \text{ has a jump in } [t_i,t_{i+1}]  \\
   
    0,& \text{otherwise},
\end{cases}
\end{equation*}
where $Y_i\added{^l}$ is $M$ dimensional $\mathcal{N}(\mu\added{_l},\delta^2\added{_l})$ and $N(t)$ is an $M \added{ \times d_N}$ dimensional homogeneous Poisson process with intensity $\lambda \added{\in \RB_+^{d_N}}$ \added{and independent components $N^l(t).$} Parameter $k$ in Equation \eqref{eq:discrete} \replaced{has $d_N$ components}{is given by} $k\added{^l} = \exp(\mu\added{_l} + \frac{\delta^2\added{_l}}{2}) - 1.$

This discretization coincides with continuous time SDEs in the Merton model described in Section \ref{sec:merton} and in the BS model when $\gamma_0=0$ \added{and Brownian motion is one dimensional.} 

To obtain a discrete stock process \added{$s$}, which we need to compute the loss function later on, we choose the initial stock value $s_0$ and set $\pi= 1$ in equation \eqref{eq:discrete}.

For option strike price $K$ we first compute option's $t_R$ claim as $F^{(j)} = (s^{(j)}_R - K)^+.$ Then we can define loss as 

\begin{equation}\label{eq:loss}
    loss(s_R,x_R) = \frac{1}{2}\frac{1}{M} \sum_{j=1}^{M} (x^{(j)}_R - F^{(j)})^2.
\end{equation}
Note that here we still, for brevity's sake, assume that the interest rate $r=0,$ since all the proceeding computations can be done with the same amount of accuracy.

\subsection{Network Architecture and Algorithm}

As mentioned above, we make use of the LSTM networks. A feed-forward network approach was also tested. However, it resulted in worse performance; hence we decided to stick with LSTM to allow more generality. We divide our learning task into two parts. First, we need to find the initial wealth value $x_0.$ This is a relatively simple task, so  a simple one-layer linear feed-forward NN with hidden dimension $d$ given by 

\begin{equation}\label{eq:linear}
L(y) = A y + b, \qquad y \in \RB^d, A \in \RB^{M \times d}, b \in \RB^M   
\end{equation}
is used.  Recall that we interpret initial wealth $x_0$ as an option price, a deterministic quantity. Hence, $x_0^j = x_0$ for each $j \in [M]$ and we may use \replaced{a vector of initial stock prices $y= [s_0, \ldots, s_0]^\top$}{$y = [1, \ldots , 1^\top]$} as an input for NN $L$. Since the wealth process in Equation \eqref{W} is a geometric jump diffusion where a positive initial value is assumed, we use the softplus activation function $f(x) = \log(1+e^x).$ 

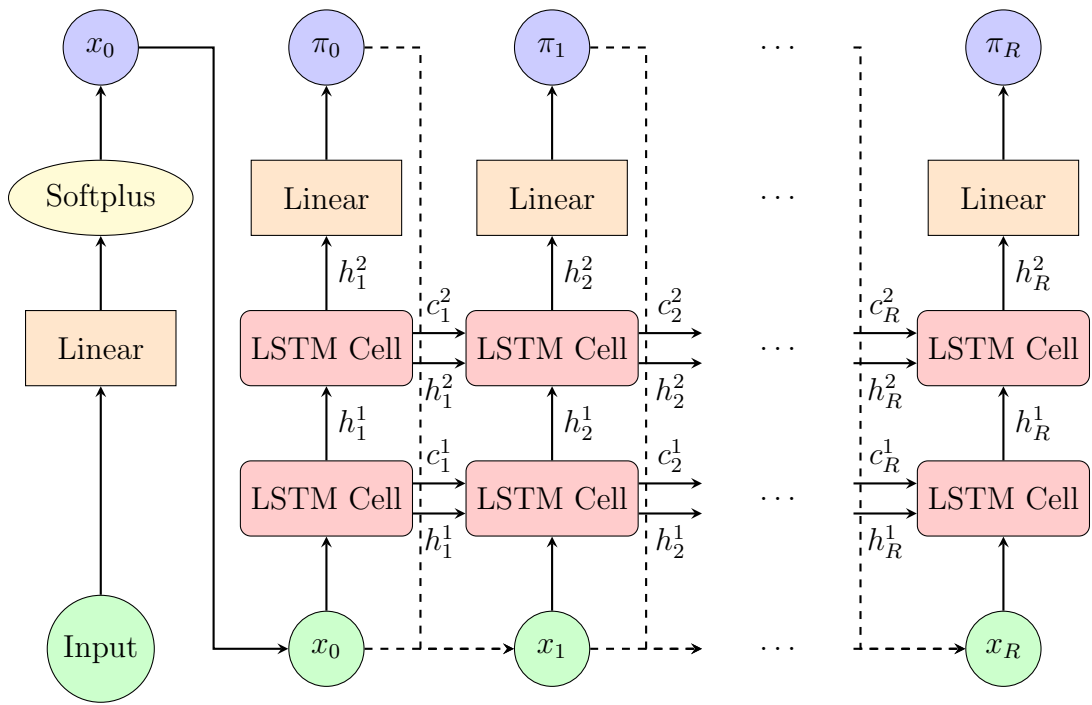
\begin{figure}[H]
    \centering
    
\begin{tikzpicture}[node distance=2cm]

\node (x0) [input] {$x_0$};
\node (lstm11) [lstm, above of=x0] {LSTM Cell};
\node (lstm12) [lstm, above of=lstm11] {LSTM Cell};
\node (nn1) [lin, above of=lstm12] {Linear};
\node (pi0) [output, above of = nn1] {$\pi_0$};

\node (input) [input, left of=x0, xshift=-1cm] {Input};
\node (nn0) [lin, left of=lstm12,xshift=-1cm] {Linear};
\node (soft) [act, left of=nn1,xshift=-1cm] {Softplus};
\node (x00) [output, left of=pi0,xshift=-1cm] {$x_0$};

\node (x1) [input, right of=x0, xshift=1cm] {$x_1$};
\node (lstm21) [lstm, above of=x1] {LSTM Cell};
\node (lstm22) [lstm, above of=lstm21] {LSTM Cell};
\node (nn2) [lin, above of=lstm22] {Linear};
\node (pi1) [output, above of = nn2] {$\pi_1$};

\node (dot1) [dot, right of=x1, xshift=1cm] {$\ldots$};
\node (dot2) [dot, above of=dot1] {$\ldots$};
\node (dot3) [dot, above of=dot2] {$\ldots$};
\node (dot4) [dot, above of=dot3] {$\ldots$};
\node (dot5) [dot, above of=dot4] {$\ldots$};

\node (xn) [input, right of=dot1, xshift=1cm] {$x_R$};
\node (lstmn1) [lstm, above of=xn] {LSTM Cell};
\node (lstmn2) [lstm, above of=lstmn1] {LSTM Cell};
\node (nnn) [lin, above of=lstmn2] {Linear};
\node (pin) [output, above of = nnn] {$\pi_R$};

\draw [arrow] (input) -- (nn0);
\draw [arrow] (nn0) -- (soft);
\draw [arrow] (soft) -- (x00);
\draw [arrow] (x00) -- ++(1.5,0) |- (x0);

\draw [arrow] (x0) -- (lstm11);
\draw [arrow] (lstm11) -- node[anchor=west] {$h_1^1$} (lstm12);
\draw [arrow] (lstm12) -- node[anchor=west] {$h_1^2$} (nn1);

\draw [arrow] ([yshift=0.2cm]lstm11.east) -- node[anchor=south] {$c_1^1$} ([yshift=0.2cm]lstm21.west);
\draw [arrow] ([yshift=-0.2cm]lstm11.east) -- node[anchor=north] {$h_1^1$} ([yshift=-0.2cm]lstm21.west);

\draw [arrow] ([yshift=0.2cm]lstm12.east) -- node[anchor=south] {$c_1^2$} ([yshift=0.2cm]lstm22.west);
\draw [arrow] ([yshift=-0.2cm]lstm12.east) -- node[anchor=north] {$h_1^2$} ([yshift=-0.2cm]lstm22.west);

\draw [arrow] (nn1) -- (pi0);
\draw [map] (pi0) -- ++(1.25,0) |- (x1);
\draw [map] (x0) --  (x1);

\draw [arrow] (x1) -- (lstm21);
\draw [arrow] (lstm21) -- node[anchor=west] {$h_2^1$} (lstm22);
\draw [arrow] (lstm22) -- node[anchor=west] {$h_2^2$} (nn2);

\draw [arrow] ([yshift=0.2cm]lstm21.east) -- node[anchor=south] {$c_2^1$} ([yshift=0.2cm]dot2.west);
\draw [arrow] ([yshift=-0.2cm]lstm21.east) -- node[anchor=north] {$h_2^1$} ([yshift=-0.2cm]dot2.west);

\draw [arrow] ([yshift=0.2cm]lstm22.east) -- node[anchor=south] {$c_2^2$} ([yshift=0.2cm]dot3.west);
\draw [arrow] ([yshift=-0.2cm]lstm22.east) -- node[anchor=north] {$h_2^2$} ([yshift=-0.2cm]dot3.west);

\draw [arrow] (nn2) -- (pi1);
\draw [map] (pi1) -- ++(1.25,0) |- (dot1);
\draw [map] (x1) --  (dot1);

\draw [arrow] ([yshift=0.2cm]dot2.east) -- node[anchor=south] {$c_R^1$} ([yshift=0.2cm]lstmn1.west);
\draw [arrow] ([yshift=-0.2cm]dot2.east) -- node[anchor=north] {$h_R^1$} ([yshift=-0.2cm]lstmn1.west);

\draw [arrow] ([yshift=0.2cm]dot3.east) -- node[anchor=south] {$c_R^2$} ([yshift=0.2cm]lstmn2.west);
\draw [arrow] ([yshift=-0.2cm]dot3.east) -- node[anchor=north] {$h_R^2$} ([yshift=-0.2cm]lstmn2.west);

\draw [map] (dot5) -- ++(1.1,0) |- (xn);
\draw [map] (dot1) --  (xn);

\draw [arrow] (xn) -- (lstmn1);
\draw [arrow] (lstmn1) -- node[anchor=west] {$h_R^1$} (lstmn2);
\draw [arrow] (lstmn2) -- node[anchor=west] {$h_R^2$} (nnn);
\draw [arrow] (nnn) -- (pin);

\end{tikzpicture}
    \caption{Neural network architecture with two LSTM layers. Dashed lines represent the update rule in Equation \eqref{eq:discrete}}
    \label{fig:nn}
\end{figure}

In the second part of the learning process, we strive to find the optimal portfolio $\pi$. We use a stacked LSTM network with two layers. Each layer is built from LSTM cells, one at each timestep. We do not give a detailed description of LSTM cells here. An interested reader can find more information in \citep*{HJ}. At each time step, the update rule \eqref{eq:discrete} is used to obtain the wealth state $x_i$, which we feed as an input for the first LSTM layer, as shown in Figure \ref{fig:nn}. It is important to note that learning the optimal portfolio $\pi$ is a challenging problem since the portfolio at each timestep depends indirectly on the previous portfolio values. Our network architecture captures this dependency through hidden states $h_i^j$ and cell states $c_i^j$ for $i \in [R]$ and $j = 1,2.$  Second LSTM layer is used mainly to take into account the non-linearity of our problem and consequently the non-linearity of the solution $\pi$. After the second LSTM layer returns its hidden state $h_i^2 \in \RB^d$, we feed it to another linear network of the form \eqref{eq:linear}, which at last yields $\pi_i$. To summarize, our network takes \replaced{as input the initial stock value $[s_0, \ldots, s_0]^\top \in \RB^d$, Brownian motion increments $B \in \RB^{M \times d_B \times R}$ and jump increments $J \in \RB^{M \times d_N \times R}$}{an input $y \in \RB^d$} and returns an output \replaced{$(x_0,\pi) \in  \RB^{M \times R+1},$}{$(x_0,\pi) \in  \RB \times \RB^N,$} where we can interpret $x_0$ as option price and $\pi$ as hedging portfolio.

Once done, we compute $loss(s_R,x_R)$ and use backpropagation to update the weights. In particular, the Adam optimizer is used. It feels befit to note here how the backpropagation approach to modifying the weights connects naturally to the theoretical control problem we solve in Section \ref{sec:optimal control}. Recall, in the Stackelberg game approach, we first determine the optimal portfolio $\hat{\pi}_z$ given arbitrary initial state $z$, and only after that do we also find the optimal initial state $\hat{z}$ where behavior under the above portfolio is assumed. Similarly, we start with an arbitrary initial state and portfolio when taking the deep learning approach. Then once the loss is computed, we modify weights in a backward manner. Hence we first modify weights responsible for the selection of portfolio $\pi_{x_0}$ given (at the moment arbitrary) initial wealth $x_0$, and only after that do we also modify the weights associated with the initial wealth.

\begin{remark}
    Since the control process $\pi$ is $\mathbb{F}$-adopted, one may feel inclined to use the Brownian motion sequence $(B_i)_{i=1}^R$ as an input for the LSTM network. Indeed, it holds  $\mathcal{F}^X_t \subseteq \mathcal{F}_t$, so it may happen that using the sequence $(x_i)_{i=1}^R$ as an input may result in sub-optimal controls. However, as it is shown in Equation \eqref{3.9a}, the optimal control is of the feedback form, hence the choice of $(x_i)_{i=1}^R$ as the data input is reasonable. Apart from that, as discussed in \citep*{han2021recurrent}, taking $(B_i)_{i=1}^R$ as an input, does not improve the performance.
\end{remark}

To summarise, let us present our Algorithm \ref{algo} in a more compact way.

\begin{algorithm}[H]
\caption{Price of a European option and a hedging portfolio in Merton model}
\label{algo}
\begin{algorithmic}
\Require Maturity time $T$, drift parameter $\alpha_0$, volatility parameter $\beta_0$, jump parameter $\gamma_0$, initial stock price $s_0$, strike price $K$, number of time steps $R$, batch size $M$ and number of epochs $P$

\For{$1\leq \text{epoch} \leq P$}
    \State Using discretization rules in Equation \eqref{eq:discrete} and NN architecture in Figure \ref{fig:nn} compute $M$ dimensional wealth sequence $(x_i)_{i \in [R]},$ stock sequence $(s_i)_{i \in [R]}$ and control sequence $(\pi_i)_{i \in [R]}$  
    \State Compute $loss(s_R,x_R)$ as in Equation \eqref{eq:loss}
    \State Using backpropagation, update the weights of the NN architecture
\EndFor
\State \Return $x_0$ and $(\pi_i)_{i \in [R]}$
\end{algorithmic}
\end{algorithm}

\subsection{Numerical Results}

Here we present the numerical results obtained using Algorithm \ref{algo}. For calculations, we used the PyTorch library from Python. \replaced{The hidden dimension is set to 512, batch size to 256, and the learning rate to 0.0005. Unless stated otherwise, we set time maturity $T=1$. To assess the results, an evaluation set of size 10000 is used. The programming code can be found on the GitHub repository: https://github.com/janrems/DeepLearningQHedging.}{The hidden dimension is set to 512, the learning step is set to 0.0005, and we set terminal time $T=1.$} \\

\replaced{We examine four market models: the standard Black-Scholes (BS) model, the BS model with multiple independent Brownian motions, the Merton model, and the Kou double exponential jump diffusion model. Our analysis focuses on the convergence of losses in all these models. For the first three models, we explore the convergence of initial values toward the specified option price as well. Additionally, we compare the predicted hedging portfolio to its analytical counterpart for both variations of the BS model. Furthermore, we assess the algorithm's performance as the input dimension increases. In the continuous case, we achieve this by comparing results from the two studied BS variations. In jump models, we test scalability concerning the input dimension in a Merton case. Finally, we investigate the scalability of the algorithm in the temporal dimension, specifically when adjusting the timestep parameter $R$ and maturity $T$.}{We take a look at three market models, namely the standard BS model, the BS model with two independent Brownian motions, and the Merton model. For all of them, we analyze the convergence of initial values towards the theoretical option price, convergence of loss function, and behaviour of the hedging portfolio.} 

\subsubsection{Black Scholes market model}\label{sec:num_BS}

\begin{figure}
\centering
\begin{subfigure}{.3\textwidth}
  \centering
  \includegraphics[width=1\linewidth]{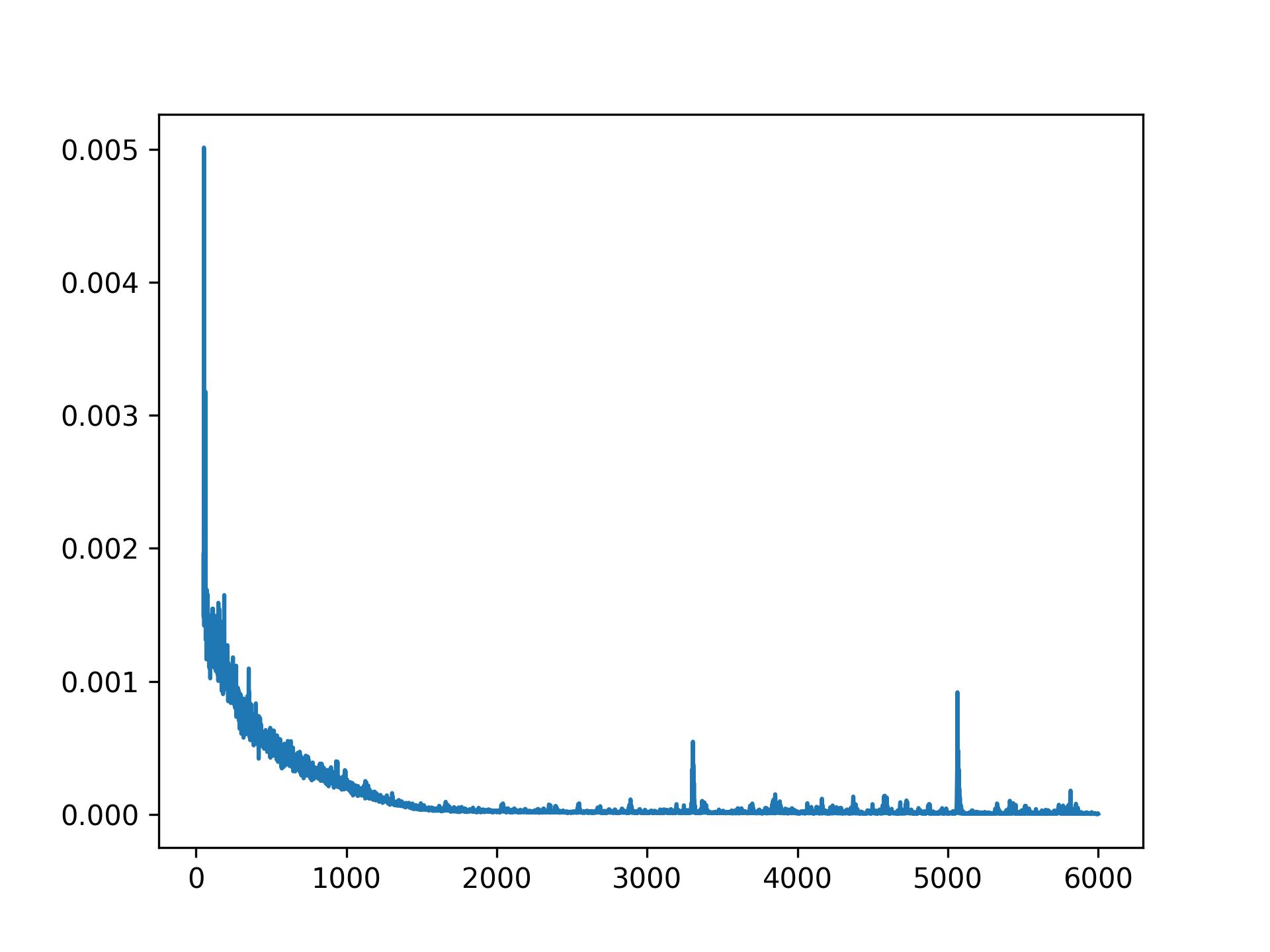}
  \caption{Loss}
  \label{graph:bs_loss}
\end{subfigure}%
\begin{subfigure}{.3\textwidth}
  \centering
  \includegraphics[width=1\linewidth]{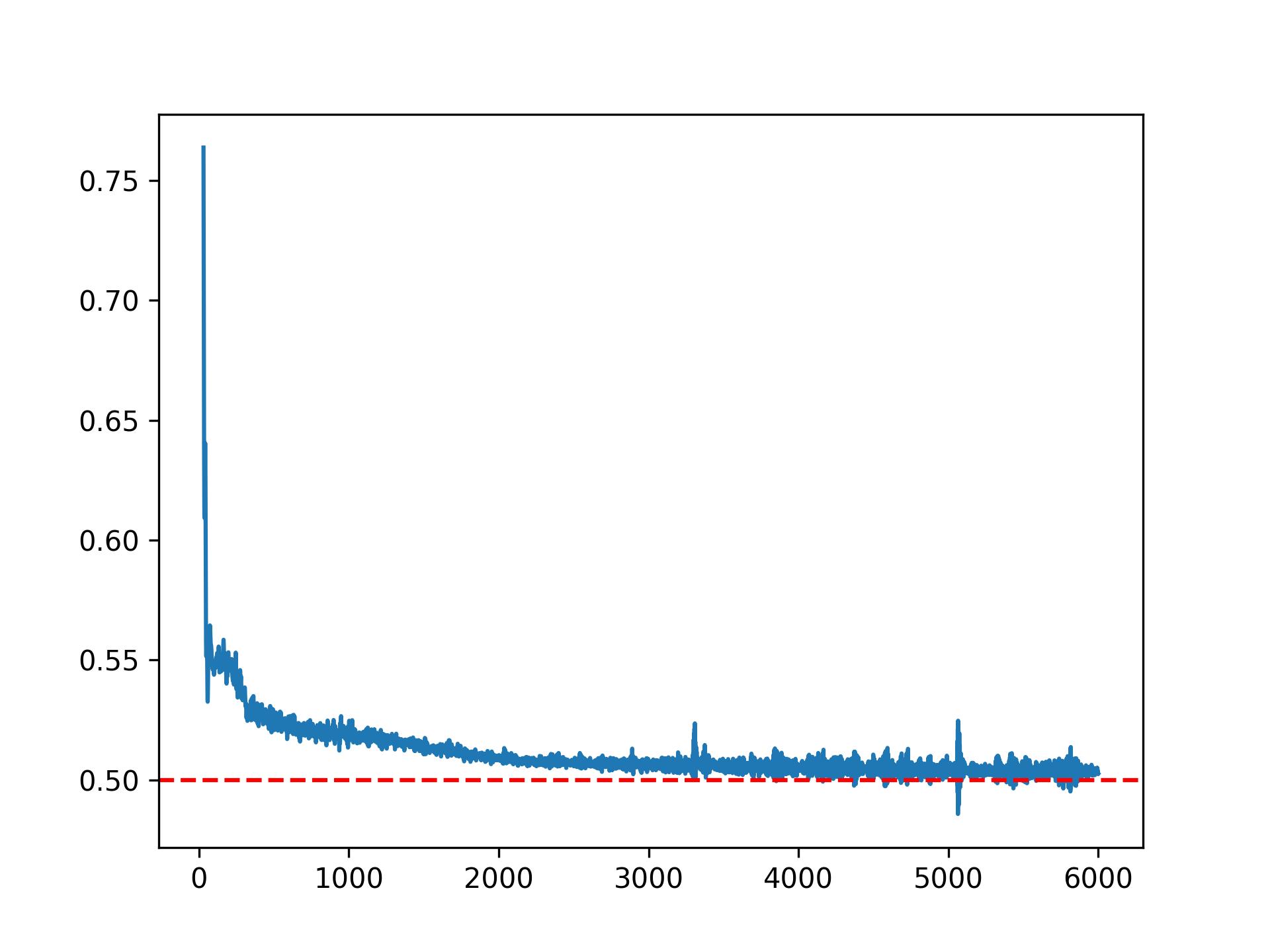}
  \caption{Initial values}
  \label{graph:bs_initial}
\end{subfigure}
\begin{subfigure}{.3\textwidth}
  \centering
  \includegraphics[width=1\linewidth]{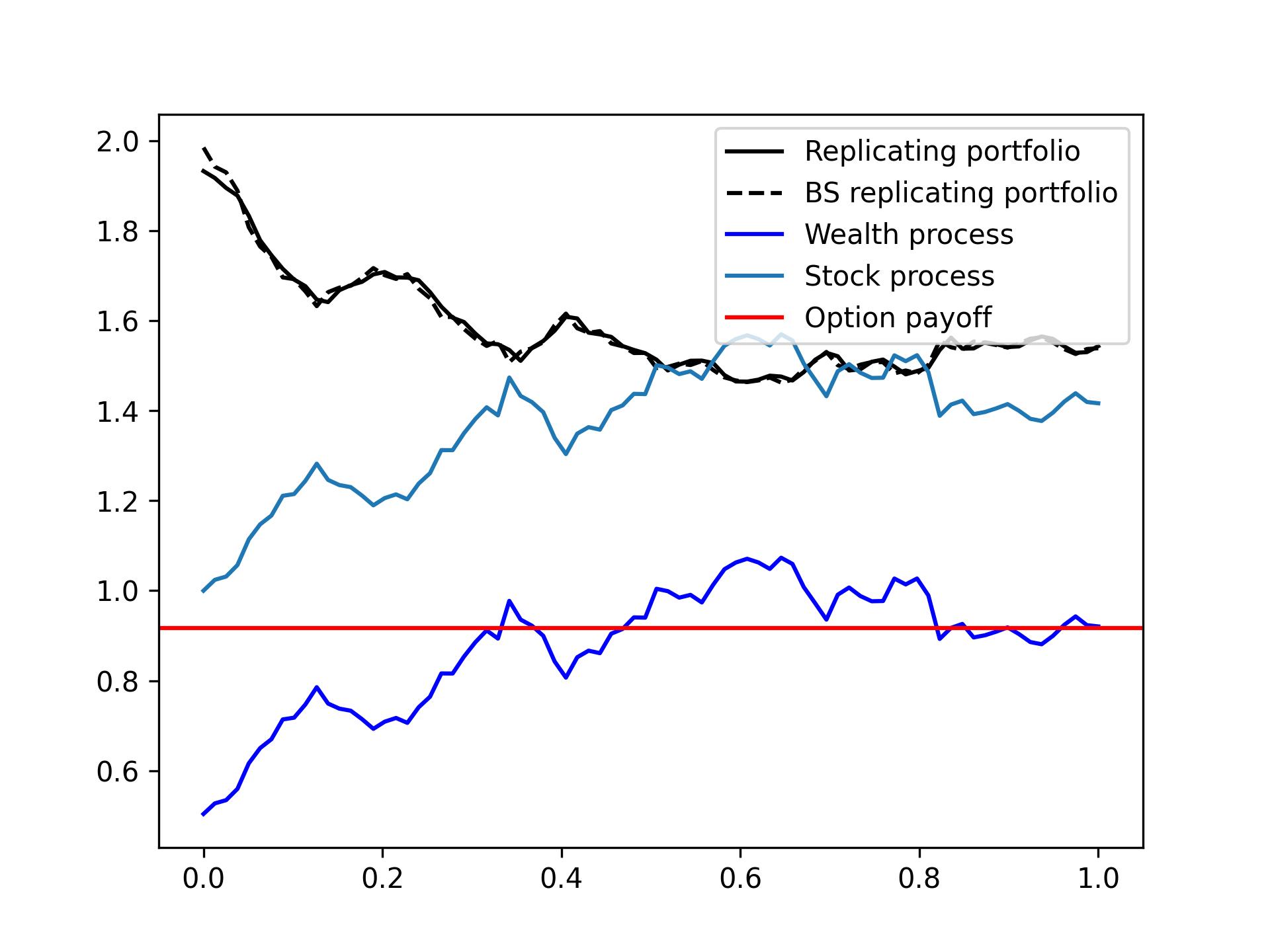}
  \caption{Market realization}
  \label{graph:marketbs}
\end{subfigure}
\caption{Convergence of loss and initial values and one market realization for the BS model}
\label{fig:testBS}
\end{figure}

For the BS model, we take the number of time steps $R=80$\deleted{ and batch size $M=256$}. Recall, the BS model is a case of the complete market model, which means that there exist initial wealth and optimal hedging portfolio that, at least in a continuous case, assures $loss = 0.$ Apart from that, we have nice analytic solutions for both option price and hedging portfolio. \\
We consider a case where in Equation \eqref{eq:discrete} we set $\alpha_0= 0.3,$ $\sigma_0=0.2,$ $\gamma_0=0$, $\lambda=0$ while we put $s_0 = 1$ and $K=0.5.$ Different choices of listed parameters result in different training times but do not affect the model's accuracy. For this particular choice of parameters, we compute 6000 epochs. \deleted{In figures \ref{graph:bs_loss} and \ref{graph:bs_initial}, we can see how loss and initial value $x_0$ both converge towards 0 and theoretical BS option price 0.5, respectively.}

\added{All the results presented here and in the other models are obtained in the following way. At the epoch where the minimal loss is achieved, we save the network weights. Then, using those weights, we evaluate our model on a larger evaluation set of size 10000.}

\replaced{The first two outputs of the algorithm are 
the optimal initial wealth $\hat{x}_0$ and the minimal obtained loss $loss_{min}$. Both of them can be found in Table \ref{table:bs}.}{As the final output of our model regarding the option price, we take the initial value that was obtained at the epoch where the minimal loss was achieved. Both optimal initial value $\hat{x}_0$ and minimal loss $loss_{min}$  can be found in Table \ref{table:bs}.} \added{In figures \ref{graph:bs_loss} and \ref{graph:bs_initial}, we can see how loss and initial value $x_0$ both converge towards 0 and theoretical BS option price 0.5, respectively.}

Now, let us take a look at our algorithm's other output, namely \added{hedging} portfolio $\pi.$ Unlike initial value $x_0$ which is deterministic, portfolio $\pi$ differs over different market realisations. \replaced{We select portfolio realizations obtained on the evaluation set and denote them}{Again, we select the portfolios at the epoch with minimal loss and denote them} with $\hat{\pi}.$ We also wish to determine how close $\hat{\pi}$ is to theoretical portfolio given by the BS model. To obtain its discrete version from our discrete stock process $s_{i \in [R]}$ we define 
\begin{equation*}
    d_i = \frac{1}{\sigma_0 \sqrt{t_R-t_i}}\Big( \log \big(\frac{s_i}{K} \big) + \frac{\sigma_0^2}{2}(t_R-t_i) \Big),
\end{equation*}
and put 
\begin{equation*}
\varphi_i = \Phi(d_i)\frac{s_i}{x_i}, 
\end{equation*}
where $\Phi$ stands for the standard normal cumulative distribution function.

We define and estimate the distance between $\hat{\pi}$ and $\varphi$ as a discrete $L^2$ distance estimate, namely

\begin{equation}\label{eq:l2}
    d(\hat{\pi}, \varphi) = E\big[ \lVert \hat{\pi} - \varphi \rVert_{L^2} \big] \approx \frac{1}{M} \sum_{j \in [M]}\sqrt{ \sum_{i \in [R]} (\hat{\pi}_i^{(j)} - \varphi_i^{(j)})^2 \Delta_t}.
\end{equation}

We can see one instance of market realization, namely the stock process $s$, state process $x$, computed portfolio $\hat{\pi}$ theoretical portfolio $\varphi$ and option payoff at terminal time $F$ in Figure \ref{graph:marketbs}.

The estimated portfolio follows the theoretical one closely. Moreover, as expected, we manage for the terminal wealth to be almost exact to the option payoff.

\begin{table}[!htb]
\centering
\begin{tabular}{ |p{1cm}||p{2cm}|p{2cm}|p{2cm}|p{2cm}|  }
 
 \hline
 Model & BS price & $\hat{x}_0$ & $ d(\hat{\pi}, \varphi)$ & $loss_{min}$  \\
 \hline
 BS   & 0.500    & \replaced{0.501}{0.5006} & \replaced{0.312}{0.018} &  \replaced{5.84e-6}{1.04e-6}\\

 \hline
\end{tabular}
\caption{Comparison of theoretical and numerical results in BS model}
\label{table:bs}
\end{table}

\FloatBarrier

\subsubsection{Black Scholes market with \replaced{multiple}{two} independent Brownian motions}

\added{In contrast to the classical Black-Scholes market with one Brownian motion, here the noise in the model comes from multiple independent Brownian options. In turn, this results in a non-complete market model. Let us recall that this means there is no guarantee that the terminal wealth $X_T$ equals the option payoff $F$ almost surely. Besides, there is, in general, a continuum of possible option prices. However, as it was discussed in Section \ref{sec:BS2B}, in the case of the European call option, following the minimal-variance hedging approach, we end up with the price and the optimal hedging strategy coming from the classical Black-Scholes market with single Brownian motion and corresponding volatility coefficient equal to the norm of volatility coefficients of multi-dimensional Brownian motion.} 

\added{This gives us a nice opportunity to study how our algorithm scales in the presence of multidimensional noisy input. We choose the same parameters as in the previous section, the only difference being the volatility coefficient, which is now, of course, multidimensional and given by $\sigma_0 = [0.11, 0.16, 0.05].$ Note that we have $\lVert \sigma_0 \rVert = 0.2$.}

\deleted{Here we use discretization slightly different than the one in Equation \eqref{eq:discrete}. The alternative discretization is of the form} 

\deleted{where $B^{(i)}_1, B^{(i)}_2$ are independent $M$-dimensional standard normal random variables. This discretization coincides with its continuous counterpart from Section \ref{sec:BS2B}. Again, discrete stock price $s$ is obtained by putting $\pi = 1.$ We use Algorithm \ref{algo2}, a slightly modified version of Algorithm \ref{algo}.}

\deleted{In this particular case we set parameters to $\beta_1 = 0.2$ and $\beta_2 = 0.3$ while we also put $\alpha_0= -0.2, s_0 = 1, K = 0.5$.}

\begin{figure}
\centering
\begin{subfigure}{.3\textwidth}
  \centering
  \includegraphics[width=1\linewidth]{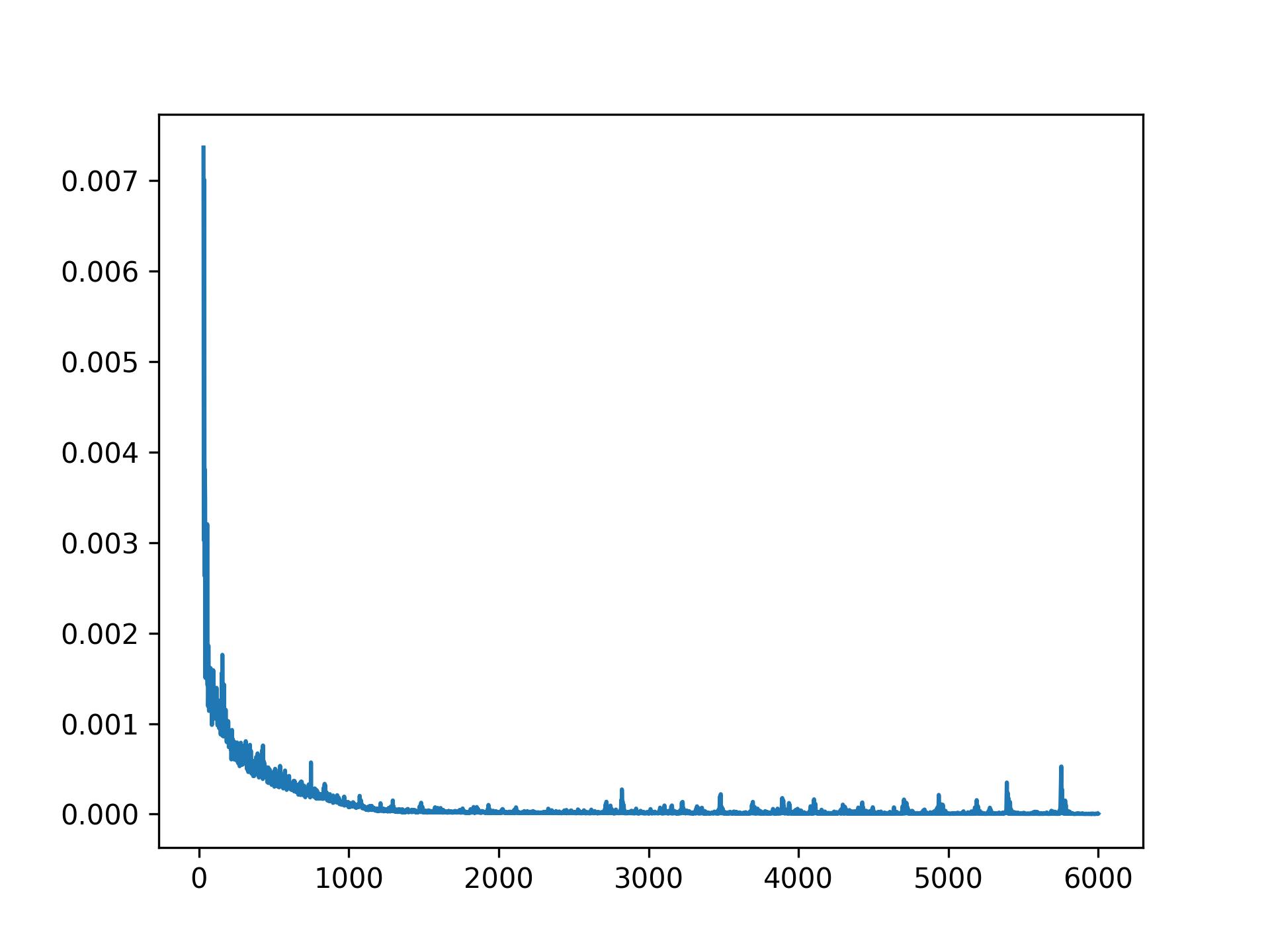}
  \caption{Loss}
  \label{graph:lossbs2}
\end{subfigure}%
\begin{subfigure}{.3\textwidth}
  \centering
  \includegraphics[width=1\linewidth]{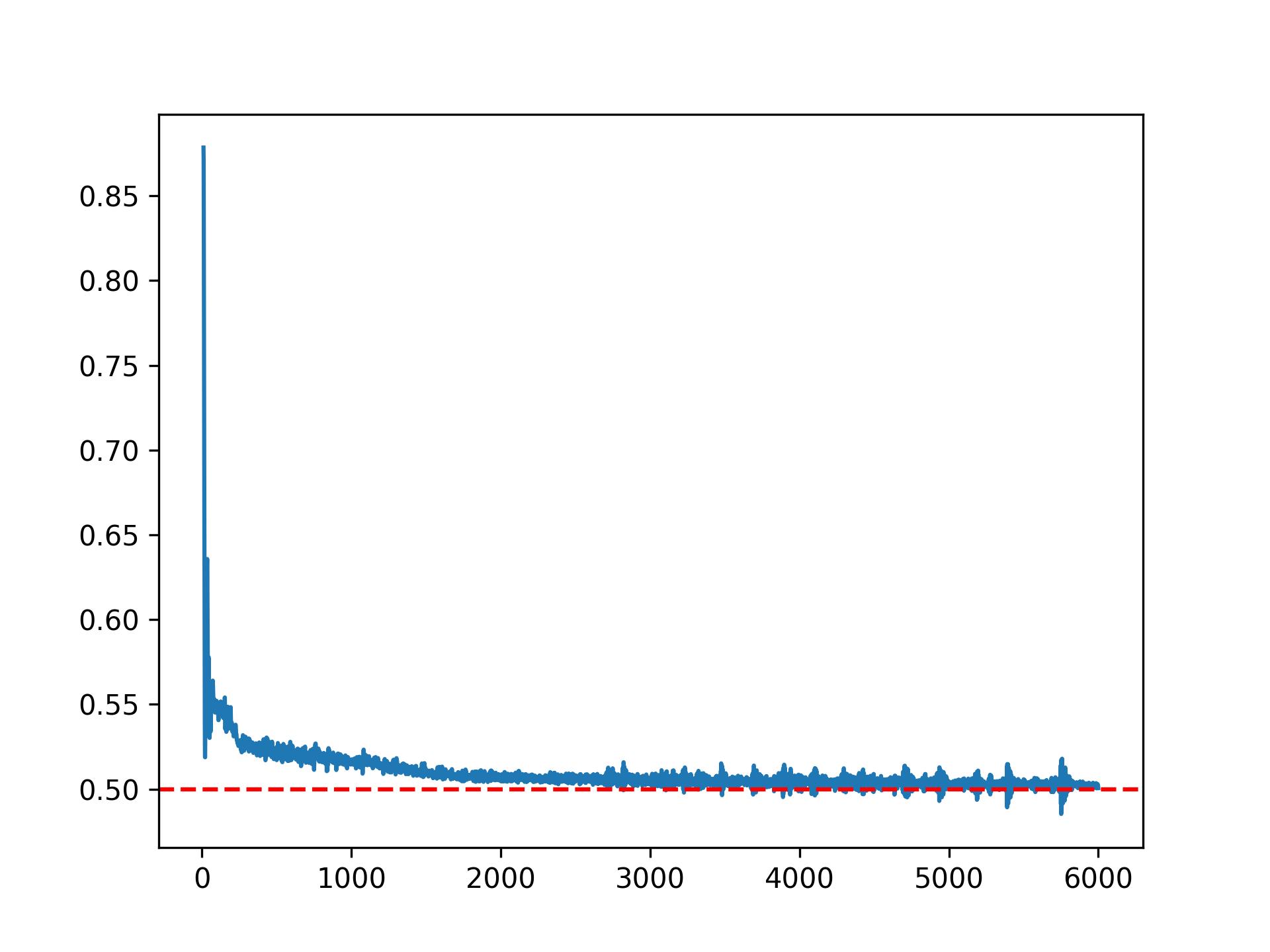}
  \caption{Initial values}
  \label{graph:initialbs2}
\end{subfigure}
\begin{subfigure}{.3\textwidth}
  \centering
  \includegraphics[width=1\linewidth]{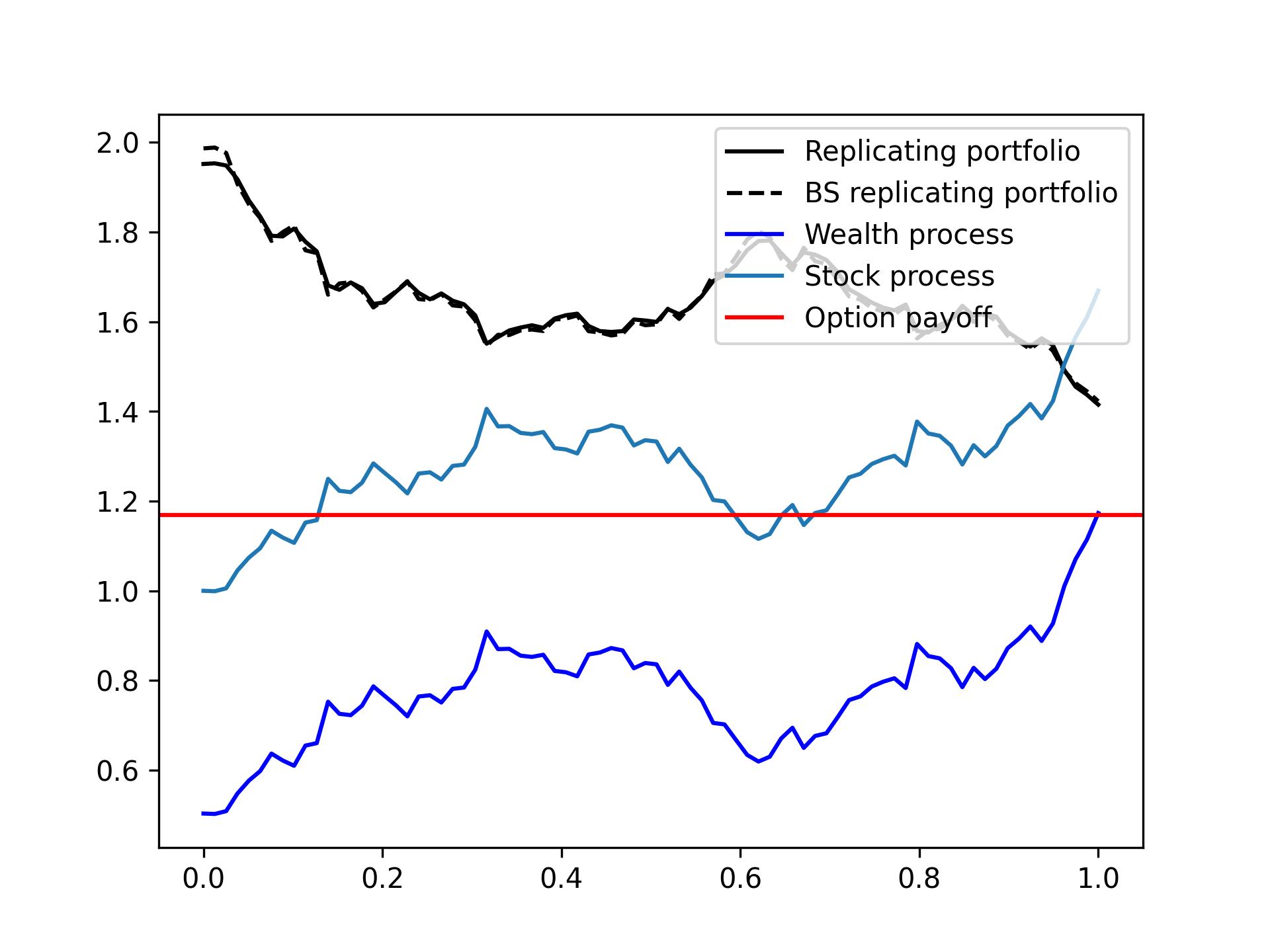}
  \caption{Market realisation}
  \label{graph:marketbs2}
\end{subfigure}
\caption{Convergence of loss and initial values and one market realization in the BS model with \replaced{multiple}{2} Brownian motions}
\label{fig:testBS2}
\end{figure}

\deleted{As in the standard BS model, we take the number of timesteps $R=30$ and batch size $M=256.$ However this type of model is known to be an incomplete market model since we have two sources of randomness. This means that the terminal wealth $X_T$ is not equal to the option payoff $F$ almost surely. Hence, we may expect the loss of our model to be higher than in the classical BS model. First, let us look at one particular realization of the market in Figure \ref{graph:marketbs2}.}

\deleted{We can see the hedging portfolio being more dynamic than in the BS model with one Brownian motion, which is expected due to increased randomness in the model.} 

\deleted{The theoretical option price for the case of this model is given by Equation \eqref{eq:bs2-price}. We estimate $E[FZ^*_T]$ using the Monte Carlo approach and obtain an estimate $\hat{z}= 0.5028$ which is only slightly higher as in the standard BS model. Convergence of the option price predicted by the deep learning algorithm and the convergence of loss can be seen in figures \ref{graph:lossbs2} and \ref{graph:initialbs2}.}

\deleted{Note that both losses and initial values appear noisier than in the previous example. In Table \ref{tab:bs2}, we can see how the option price predicted by our algorithm compares to the one obtained with the Monte-Carlo approach and what is the minimal obtained loss. }

\added{Comparing both results in Table \ref{tab:bs2} and graphs in Figure \ref{fig:testBS2} with their counterparts in Section \ref{sec:num_BS}, we can see that our algorithm scales well with the increased size of the input. Even though the results in both tables differ a bit, it is only natural because of the random nature of our machine learning algorithm.}

\begin{table}[!htb]
\centering
\begin{tabular}{ |p{2cm}||p{2cm}|p{2cm}|p{2cm}|p{2cm}| }
 \hline
 Model & \replaced{BS price}{$\hat{z}$} & $\hat{x}_0$ & \added{$ d(\hat{\pi}, \varphi)$} &$loss_{min}$  \\
 \hline
 \replaced{BSMB}{BS2B}   & \replaced{0.500}{0.5028}    &\replaced{0.502}{0.5038} & \added{0.226}& \replaced{6.14e-6}{$3.52e-5$}\\
 \hline
\end{tabular}
\caption{Comparison of theoretical and numerical results for the BS model with \replaced{multiple}{2} Brownian motions (\replaced{BSMB}{BS2B})}
\label{tab:bs2}
\end{table}

\subsubsection{Merton market model}\label{sec:num_merton}

In our particular example of the Merton model, we follow the setting from Section \ref{sec:merton} and again use Algorithm \ref{algo}, where we set parameters to $\alpha_0= 0.2,$ $\sigma_0=0.2,$ $\gamma_0=1,$ $s_0 = 1$ and $K=0.5.$ For the log-normally distributed jumps, we set expected value $\mu = -0.2$ and standard deviation $\delta = 0.05,$ while jump intensity $\lambda=5.$ Because of the downward jumps, we need to work with finer discretization to avoid negative wealth and stock values. Hence, we choose $R=150$, which in turn leads to longer training time since we now have 150 times 2 LSTM cells in our NN. As the stock process is noisier than the one in the BS case, so is the portfolio, which in turn also means a harder learning task. We use 7,000 epochs \added{and disregard the training rounds where negative jumps occur}. 

\begin{figure}[H]
\centering
\begin{subfigure}{.3\textwidth}
  \centering
  \includegraphics[width=1\linewidth]{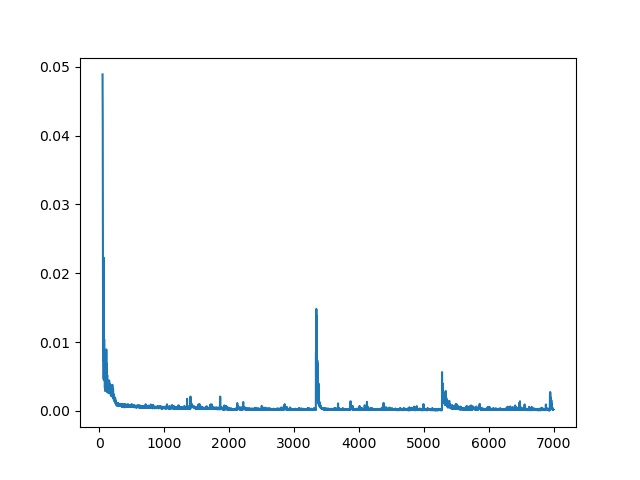}
  \caption{Loss}
  \label{graph:lossm}
\end{subfigure}%
\begin{subfigure}{.3\textwidth}
  \centering
  \includegraphics[width=1\linewidth]{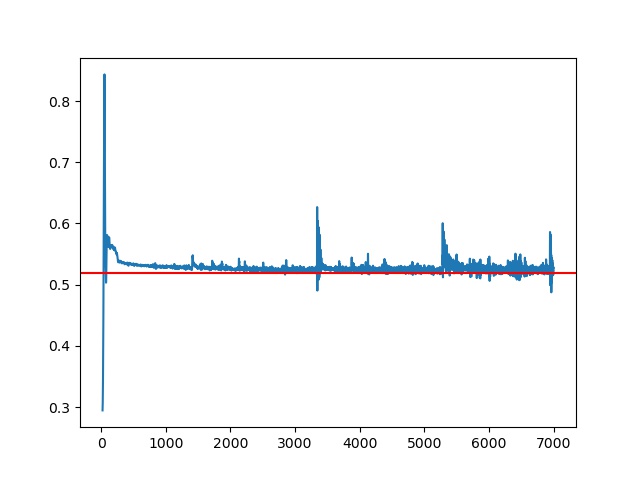}
  \caption{Initial values}
  \label{graph:initialm}
\end{subfigure}
\begin{subfigure}{.3\textwidth}
  \centering
  \includegraphics[width=1\linewidth]{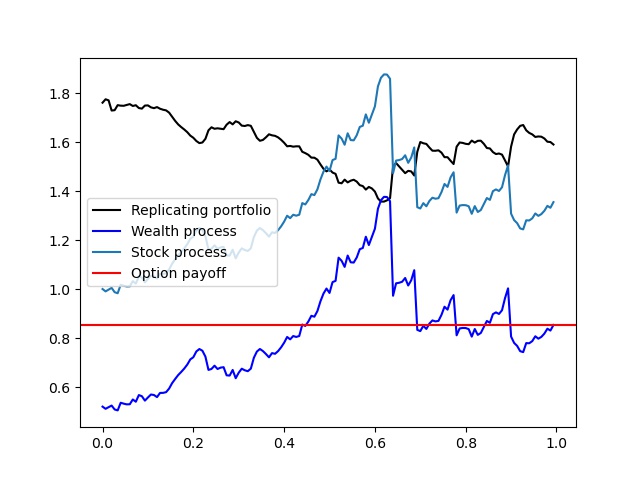}
  \caption{Market realization}
  \label{graph:marketm}
\end{subfigure}

\caption{Convergence of loss and initial values and one market realization for the Merton model}
\label{fig:testM}
\end{figure}

In Figure \ref{graph:marketm}, we see one market realization at the epoch at which the lowest loss was obtained.

The portfolio is much more dynamic, which, as discussed, results in slower learning. Furthermore, we see a more significant gap between terminal wealth and option payoff.

As we can see in the figures \ref{graph:lossm} and \ref{graph:initialm}, both losses and initial values converge toward desired quantities. However, we can see that convergence becomes slow after just 3000 epochs. Tackling the plateau problem with, for instance, scheduling the learning rate is one of the tasks for future work. Notice there are a few spikes in both of these graphs. We believe they are due to computational errors that occur when the wealth process takes values close to zero, which is more often than in the previous two cases due to downward jumps.

\begin{figure}[H]
\centering
\includegraphics[width=0.5\linewidth]{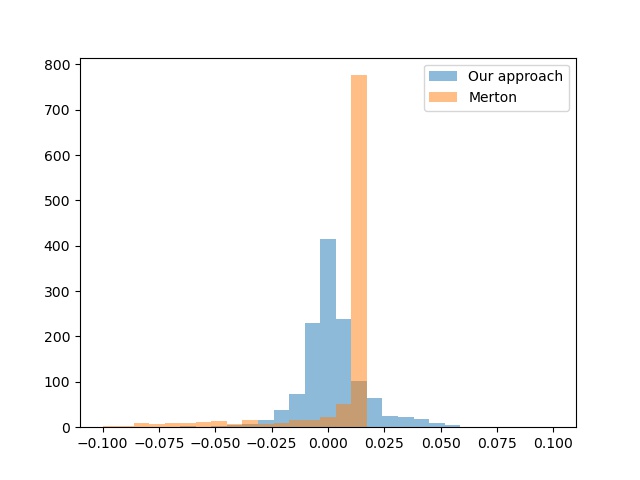}

\caption{Distribution of the difference between terminal wealth and option payoff under Merton's and our approach in 1000 realizations}
\label{fig:lossdist}
\end{figure}

The option price for the case of the Merton model is given by Equation \eqref{eq:merton price}. Using the Monte Carlo approach, we get an estimation $\hat{z} = 0.519$. Here, we again emphasize that this price differs from the one presented by Merton in Equation \eqref{eq:merton}, which equals 0.515 for our particular choice of parameters. This means that the difference between the option price proposed in \citep*{M} and the minimal variance price is nontrivial in practice. In Figure \ref{fig:lossdist}, we can see how the distribution of the differences between terminal wealth and option payoff in Merton's and our approach differ. We can see that using Merton's portfolio, one usually observes small gains, which are balanced with occasional large losses. On the other hand, the distribution is symmetric around zero using our approach. Apart from that, we are able to avoid large losses entirely, which is an advantage. Consequently, using a minimal variance hedging strategy is more costly, as discussed in Section \ref{sec:merton}.

The performance of our deep learning algorithm for the Merton model is presented in Table \ref{tab:merton}. \replaced{Compared to the performance in the BS case, the error of price prediction is around ten times higher, which is due to the increased volatility in the model.}{Compared to the performance in the case of a continuous incomplete market, the error of price prediction is around two times higher.}
\begin{table}[!htb]
\centering
\begin{tabular}{ |p{2cm}||p{2cm}|p{2cm}|p{2cm}| }
 
 \hline
 Model & $\hat{z}$ & $\hat{x}_0$ &  $loss_{min}$  \\
 \hline
 Merton   & 0.519    &0.521& 8.13e-5\\
 \hline
\end{tabular}
\caption{Comparison of theoretical and numerical results in Merton model}
\label{tab:merton}
\end{table}

In the case of both complete and incomplete markets, we showed that the deep learning algorithm performs well by comparing the results with cases where explicit solutions or at least Monte Carlo estimates are available. 

\added{To analyze the scalability of the algorithm concerning an increase in input dimension, similar to the continuous case, we examine the Merton model jump diffusion process driven by a multi-dimensional compound Poisson process. Specifically, we compare the outcomes with a one-dimensional version guided by a modified compensated Poisson process using mixed distributions. We make use of the following well-known result.}

\begin{proposition}
    \added{Let $X_1(t) = \sum_{i=1}^{N_1(t)} Y^1_i$ and $X_2(t) = \sum_{i=1}^{N_2(t)} Y^2_i$ be two compound Poisson processes where $N_1(t)$ and $N_2(t)$ are independent homogenous Poisson processes with intensities $\lambda_1$ and $\lambda_2$ respectively. Furthermore, jumps $Y_i^1$ and $Y_i^2$ are independent, coming from possibly different distributions. Consider now $N(t)$, a homogenous Poisson process with intensity $\lambda = \lambda_1 + \lambda_2$ and random variables}
    \begin{align*}
        Y_i \sim \begin{pmatrix}
    &Y_i^1  \qquad &Y_i^2 \\
    &\frac{\lambda_1}{\lambda_1 + \lambda_2} \qquad &\frac{\lambda_2}{\lambda_1 + \lambda_2}
  \end{pmatrix}.
    \end{align*}
    \added{Then processes $X(t) = \sum_{i=1}^{N(t)} Y_i$ and $X_1(t) + X_2(t)$ have the same distribution.} 
\end{proposition}

\added{This result easily extends to the sum of an arbitrary number of compensated Poisson processes. For simulation purposes, we select a three-dimensional compound Poisson process with intensities $[3, 5, 2]$. The lognormal jumps have means $[0.1, 0.1, 0.05]$ and standard deviations $[0.05, 0.02, 0.01]$. The obtained results are presented in Table \ref{tab:merton-mixed}. Once again, the similarity in the results indicates that our algorithm scales effectively even in cases involving discontinuous multi-dimensional input.}

\begin{table}[!htb]
\centering
\begin{tabular}{ |p{3cm}||p{2cm}|p{2cm}|p{2cm}| }
 
 \hline
 Model & $\hat{z}$ & $\hat{x}_0$ &  $loss_{min}$  \\
 \hline
 Merton 3-D   & 0.507    &0.506& 7.01e-5\\
 \hline
 Merton mixed   & 0.507    &0.506& 7.54e-5\\
 \hline
\end{tabular}
\caption{Comparison of theoretical and numerical results in Merton model}
\label{tab:merton-mixed}
\end{table}

\subsubsection{\added{Kou market model}}

\added{Here, we present numerical results for a market model introduced in \citep*{kou2002jump}. Kou proposed a model where the stock's behaviour is governed by a jump diffusion process where the jumps have a double exponential distribution. This means we can use the same type of discretization as for the Merton case described in equations \eqref{eq:discrete} and \eqref{eq:J_i}, where the jump variables $Y_i^l$ now follow a double exponential distribution}

    \begin{align*}
        Y_i^l \sim \begin{pmatrix}
    &\text{Exp}(\eta^l_1)  \qquad & -\text{Exp}(\eta^l_2) \\
    &p^l \qquad &1-p^l
  \end{pmatrix},
    \end{align*}

\added{where $p^l \in [0,1]$ represents the probability of exponentially distributed jump being either upward or downward. To ensure the existence of moments, we need additional conditions on intensities of the positive jump part, hence $\eta^l_1 >1$ while $\eta_2^l > 0$. The compensator of the drift in Equation \eqref{eq:discrete} changes accordingly and has components}
$$k^l = p^l \frac{\eta^l_1}{\eta^l_1 - 1} + (1-p^l) \frac{\eta_2^l}{\eta_2^l + 1} -1.$$

\added{Unfortunately, the property of jumps being both positive and negative and having heavier tails makes the Kou model impractical when one wants to obtain option price through equivalent martingale measures and Mont-Carlo approach as described in Section \ref{sec:merton}. The reason is that the condition $G(e^{Y_i^l}-1) >-1$, for $G$ as in Equation \eqref{eq:G}, is not satisfied for general choice of parameters $\eta_1^l$, $\eta_2^l$ and $p^l$.}

\begin{figure}
\centering
\begin{subfigure}{.3\textwidth}
  \centering
  \includegraphics[width=1\linewidth]{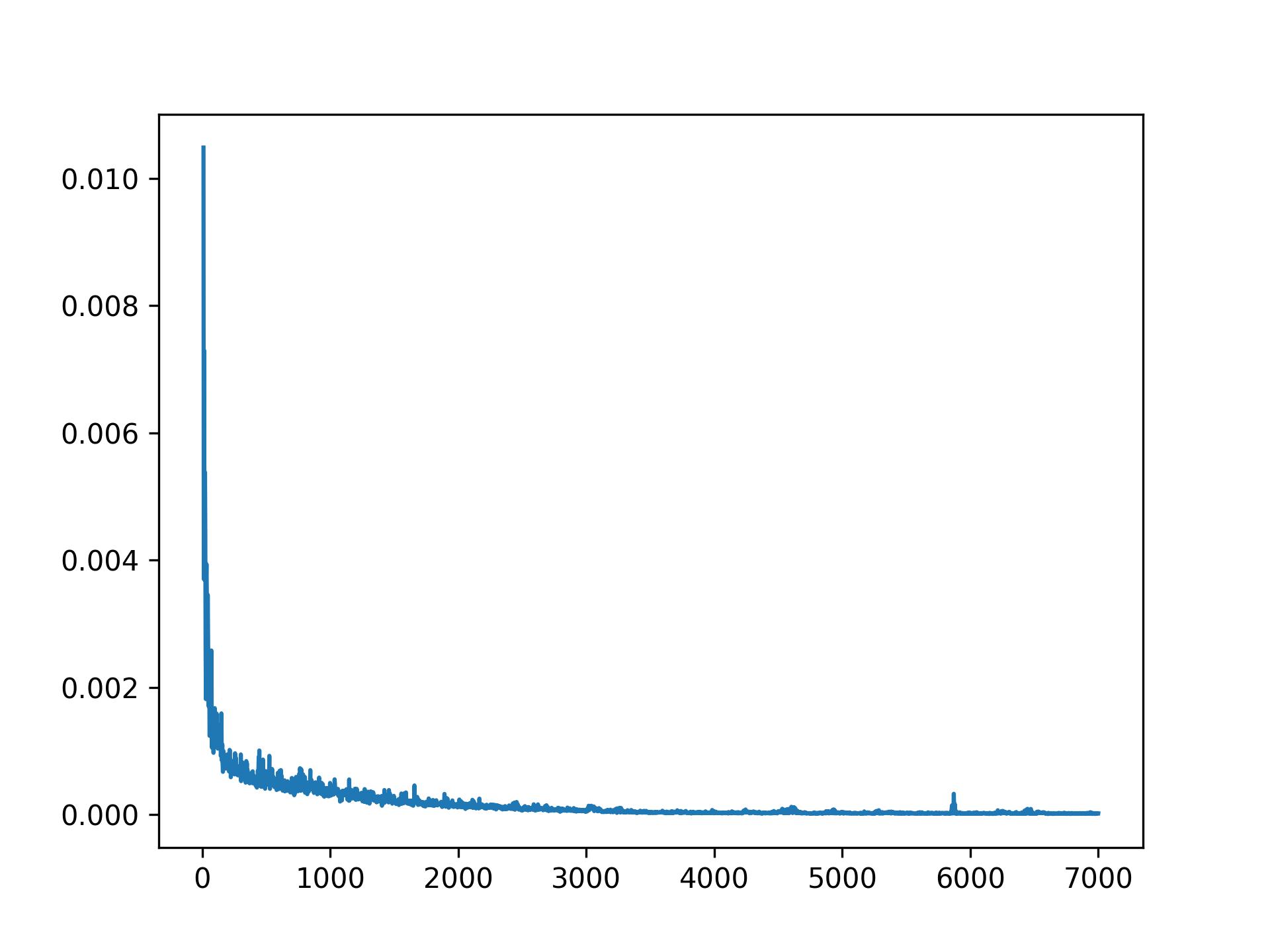}
  \caption{Loss}
  \label{graph:kou_loss}
\end{subfigure}%
\begin{subfigure}{.3\textwidth}
  \centering
  \includegraphics[width=1\linewidth]{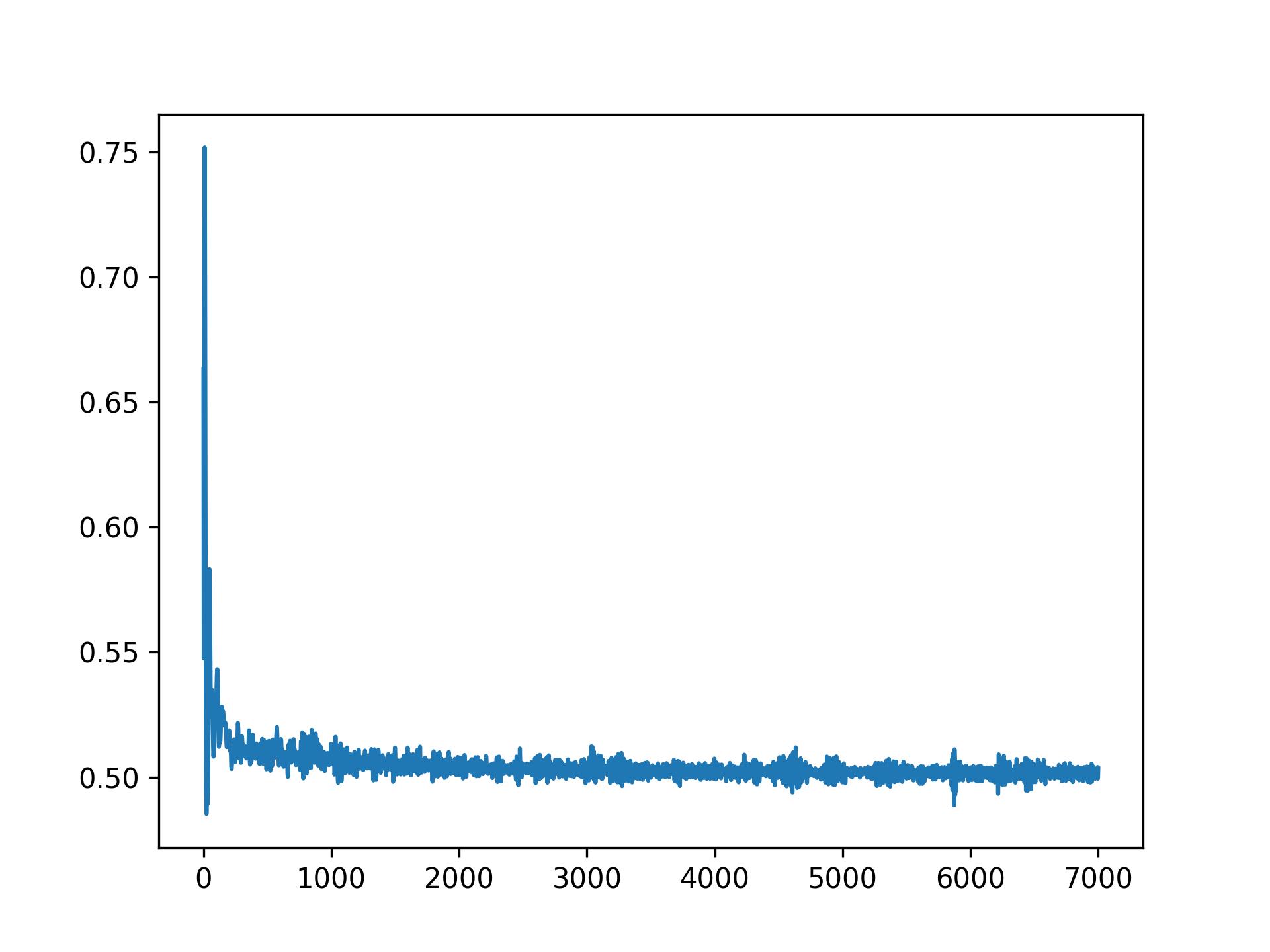}
  \caption{Initial values}
  \label{graph:kou_initial}
\end{subfigure}
\begin{subfigure}{.3\textwidth}
  \centering
  \includegraphics[width=1\linewidth]{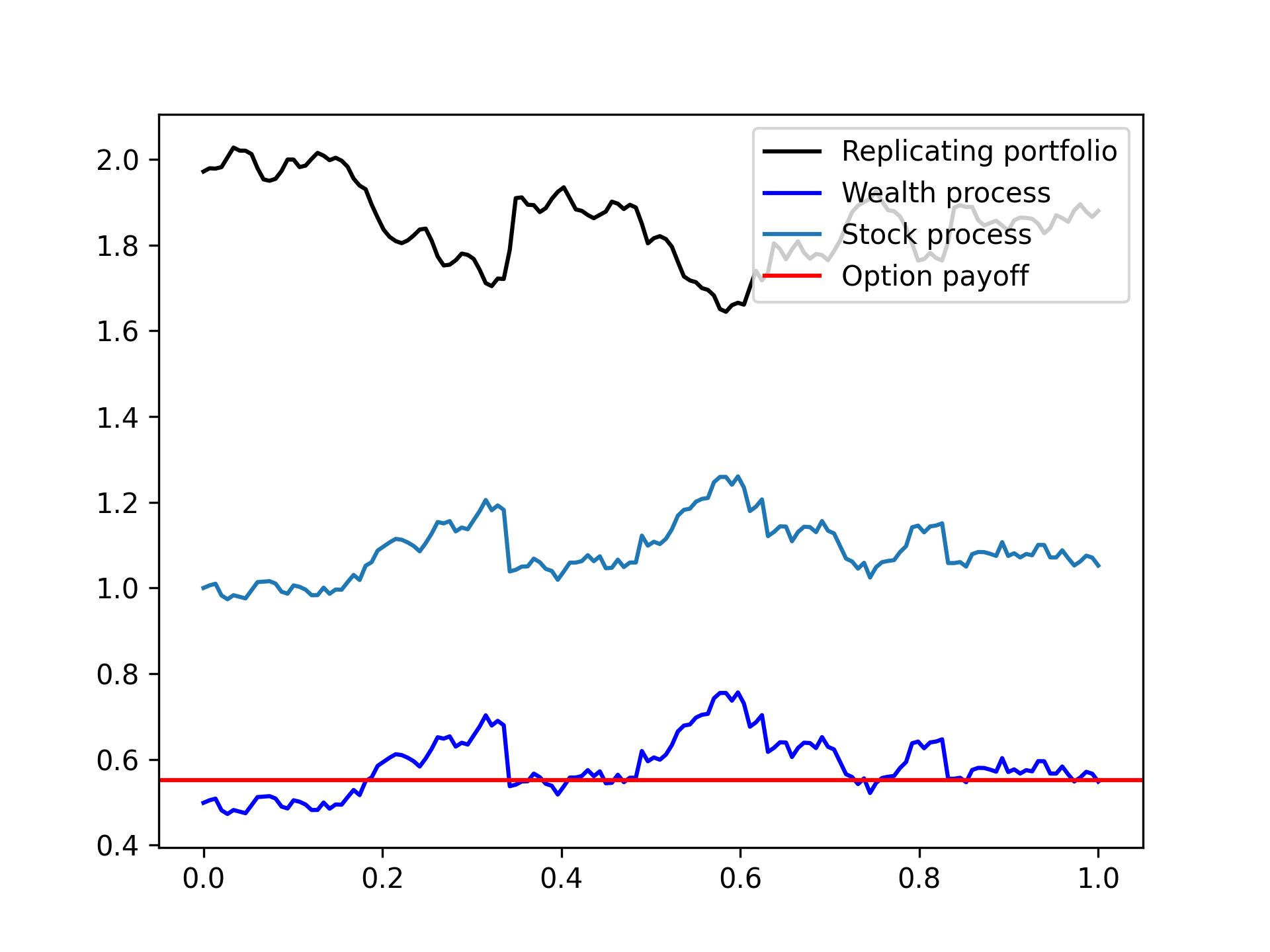}
  \caption{Market realization}
  \label{graph:marketkou}
\end{subfigure}
\caption{Convergence of loss and initial values and one market realization for the Kou model}
\label{fig:fig_kou}
\end{figure}

\added{Nonetheless, the option price can still be obtained through the use of the deep learning algorithm. Again, to avoid negative wealth and stock values, we select $R=150.$ Due to the heaviness of the tails in jump distribution, one needs to be careful when choosing parameters. In particular, $\eta_2$ shouldn't be too small. Here we choose $s_0 = 1, K=0.5,$ $ \alpha_0 = 0.15, \sigma_0 = 0.2, \gamma_0 = 1, \lambda=10, \eta_1 = 50, \eta_2 = 25$ and $p=0.3$. According to \citep*{kou2002jump}, these parameters should reflect the ones observed on the US stock market. Graphs of the loss and initial value convergence and a graph of one market realization can be found in Figure \ref{fig:fig_kou}. The minimal obtained loss is 1.54e-5, lower than the losses recorded for the Merton model due to the choice of less noisy parameters. The algorithm predicts the initial wealth value $x_0 = 0.4987$. We know that the true option price should be higher than the BS price of 0.5 but not by much due to more conservative jump parameters. Hence, the predicted result does not seem too far off. The reason for the undershoot in the predicted initial value is the inclusion of the positive jumps in the Kou model.}

\added{A possible additional direction for advancing the current study is to explore the utilization of deep learning algorithms in pricing American and path-dependent options within the context of the double exponential jump diffusion model, drawing inspiration from the research conducted in \citep*{kou2004option}.}

\subsubsection{\added{Scalability in temporal dimension}}
\added{In this section, we explore how the algorithm's performance is influenced by variations in the temporal dimension. Specifically, we consider different time maturities $T$ and varying numbers of timesteps $R$.} 

\added{We examined the algorithm's performance within the context of both the complete BS market model and the incomplete Merton market model. Various combinations of parameters $T$ and $R$ were scrutinized. The financial model parameters remain consistent with those outlined in Sections \ref{sec:num_BS} and \ref{sec:num_merton}.}

\added{Concerning machine learning parameters, we maintained a batch size of 256, a hidden dimension of 512, and a learning rate of 0.0005. For practical considerations, we conducted the learning process for a fixed 3000 epochs in each studied case. In both market models, the algorithm's performance was evaluated by comparing the loss and the absolute error of predicted option prices. For the BS model, the estimated expected $L^2$ loss defined in Equation \eqref{eq:l2} was also included in the assessment.}

\added{Results are presented in Tables \ref{tab:BS_scale} and \ref{tab:merton_scale} for the BS and Merton models, respectively.}

\begin{table}[ht]
    \centering
    \begin{subtable}{\linewidth}
        \centering
        \begin{tabular}{|*{4}{>{\centering\arraybackslash}p{1.5cm}|}}
            \hline
            \textbf{R/T} & \textbf{0.5} & \textbf{1.0} & \textbf{2.0} \\
            \hline
            \textbf{40} & 6.058e-5   & 2.917e-5   & 8.198e-5   \\
            \hline
            \textbf{80} & 4.563e-5   & 1.716e-5   & 8.246e-5   \\
            \hline
            \textbf{160} & 4.122e-5   & 2.164e-5   &  3.41e-5  \\
            \hline
        \end{tabular}
        \caption{Loss}
        \label{tab:bs_loss}
    \end{subtable}%
    
    \vspace{0.5cm} 
    
    \begin{subtable}{\linewidth}
        \centering
        \begin{tabular}{|*{4}{>{\centering\arraybackslash}p{1.5cm}|}}
            \hline
            \textbf{R/T} & \textbf{0.5} & \textbf{1.0} & \textbf{2.0} \\
            \hline
            \textbf{40} &0.0026 &0.0103 &0.0323 \\
            \hline
            \textbf{80} & 0.0020&0.0094 &0.0329 \\
            \hline
            \textbf{160} & 1.580e-5   & 0.0080 & 0.0286 \\
            \hline
        \end{tabular}
        \caption{Absolute error of option price}
        \label{tab:bs_err}
    \end{subtable}%
    
    \vspace{0.5cm} 
    
    \begin{subtable}{\linewidth}
        \centering
        \begin{tabular}{|*{4}{>{\centering\arraybackslash}p{1.5cm}|}}
            \hline
            \textbf{R/T} & \textbf{0.5} & \textbf{1.0} & \textbf{2.0} \\
            \hline
            \textbf{40} &0.3171 &0.4369 &0.7221 \\
            \hline
            \textbf{80} & 0.4161&0.5951 &1.1797 \\
            \hline
            \textbf{160} &0.4399 &1.0670 & 1.3513 \\
            \hline
        \end{tabular}
        \caption{Expected $L^2$ distance}
        \label{tab:bs_l2}
    \end{subtable}
    
    \caption{\added{Comparison of algorithm performance for different choices of timesteps $R$ and maturities $T$ - BS model}}
    \label{tab:BS_scale}
\end{table}

\added{It should be stated that due to the random nature of both inputs and the algorithm presented results are noisy in nature as well.}

\added{Nonetheless, we may with large certainty conclude that the algorithm's performance decreases with increasing maturity. This is particularly evident in the absolute error of option prices (Table \ref{tab:bs_err}) and the expected $L^2$ distance (Table \ref{tab:bs_l2}). The use of discrete Euler-Maruyama approximation to the continuous solution likely contributes to this trend, where errors accumulate over time. We believe enhanced computing power may contribute to improved performance through finer discretization, extended training time, and higher hidden dimensions.}

\added{In examining performance concerning the discretization parameter $R$, it is harder to draw strong conclusions since the difference in results is less significant. It seems that finer discretization positively impacts the algorithm's performance in terms of loss and absolute error in option prices. Conversely, an inverse trend is observed in the expected $L^2$ distance. This discrepancy may arise from increased variation in the discrete theoretical portfolio as the number of timesteps grows. Nevertheless, focusing on loss and option price, which are the key outputs of our algorithm, we find a positive response to an increased number of discretization points. As discussed earlier, this may prove beneficial when considering longer maturities.}

\begin{table}[ht]
    \centering
    \begin{subtable}{\linewidth}
        \centering
        \begin{tabular}{|*{4}{>{\centering\arraybackslash}p{1.5cm}|}}
            \hline
            \textbf{R/T} & \textbf{0.5} & \textbf{1.0} & \textbf{2.0} \\
            \hline
            \textbf{40} & 1.976e-5   & 8.527e-5   &0.0003 \\
            \hline
            \textbf{80} & 3.156e-5   & 4.914e-5   &0.0002 \\
            \hline
            \textbf{160} &  1.482e-5  &  3.728e-5  & 0.0002\\
            \hline
        \end{tabular}
        \caption{Loss}
        \label{tab:merton_loss}
    \end{subtable}%
    
    \vspace{0.5cm} 
    
    \begin{subtable}{\linewidth}
        \centering
        \begin{tabular}{|*{4}{>{\centering\arraybackslash}p{1.5cm}|}}
            \hline
            \textbf{R/T} & \textbf{0.5} & \textbf{1.0} & \textbf{2.0} \\
            \hline
            \textbf{40} & 0.0004& 0.0085&0.0290 \\
            \hline
            \textbf{80} &0.0011 & 0.0072& 0.0150 \\
            \hline
            \textbf{160} &0.0024 &0.0063 & 0.0127\\
            \hline
        \end{tabular}
        \caption{Option price absolute error}
        \label{tab:merton_err}
    \end{subtable}%
    
    \caption{\added{Comparison of algorithm performance for different choices of timesteps $R$ and maturities $T$ - Merton model}}
    \label{tab:merton_scale}
\end{table}

\added{Before presenting the results for the Merton case, we must first address the problem regarding the use of discretizations with a small number of timesteps, which was already mentioned in Section \ref{sec:num_merton}. To circumvent the problem of negative wealth and stock values, we present an alternative discretization approach. Instead of directly working with the discrete wealth process $x$ given by the update rule in Equation \eqref{eq:discrete}, we take its logarithm. Using It\^{o}'s formula and Euler-Maruyama scheme we obtain its discrete version through an update rule} 
    \begin{align*}
       \added{y_{i+1} = y_i + [(\alpha_0 - \lambda^\top k) \pi_t - \frac{\pi_t^2 \sigma_0^\top \sigma_0 }{2}] \Delta_t + \pi_t  \sqrt{\Delta_t} B_i \sigma_0+ \log( \gamma_0 \pi_t J_i +1),} 
    \end{align*}
\added{where $y_0 = \log(x_0)$ and then put $x_i = \exp(y_i)$. The same is done for the discrete stock process $s$ where we additionally put $\pi = 1$.}

\added{Let us now present the results regarding the Merton case. In tables \ref{tab:merton_loss} and \ref{tab:merton_err}, we see that the algorithm responds to changes in maturity and the number of timesteps similarly as in the BS case. The algorithm's worse performance is due to the increased randomness in the model, as it was explained in Section \ref{sec:num_merton}. Additionally, it must be mentioned that the algorithm's performance slightly decreases when the logarithmic approach outlined above is used. This decline could be attributed to the non-linear dependence of the process $y$ on the control variable $\pi$. Hence, for the jump diffusion models, we advise the use of the original update rule with the number of timesteps at least 150.} 

\section*{Data availability}

The author confirms that all data generated or analysed during this study are included in this published article. Furthermore, primary and secondary sources and data supporting the findings of this study were all publicly available at the time of submission.

\bibliography{references}

\section*{Competing interests}
The author declares no competing interests.

\end{document}